\documentclass[twocolumn,showpacs,aps]{revtex4}
\usepackage{graphicx}
\usepackage{xcolor}
\usepackage{subfigure}
\def\scr{\rm\scriptscriptstyle }
\usepackage{amssymb}
\usepackage{graphicx}
\usepackage{subfigure}
\usepackage{amsmath}
\usepackage[mathscr]{eucal}
\usepackage{lipsum} 
\usepackage{ulem}
\def\scs{\scriptscriptstyle }

\def\scr{\rm\scriptscriptstyle }

\begin{document}
\title{	Complete and incomplete fusion of $^{7}$Li projectiles on heavy targets}

\author{M. R. Cortes} 
	\email{mariane.cortess@gmail.com}	
	\affiliation{Instituto de F\'{\i}sica, Universidade Federal Fluminense, Av. Litoranea s/n, Gragoat\'{a}, Niter\'{o}i, RJ, 24210-340, Brazil}	
\author{J. Rangel} 
\email{jeannierangel@gmail.com}
	\affiliation{Instituto de F\'{\i}sica, Universidade Federal Fluminense, Av. Litoranea s/n, Gragoat\'{a}, Niter\'{o}i, RJ, 24210-340, Brazil}
\author{J. L. Ferreira} 
\email{jonas@if.uff.br}
	\affiliation{Instituto de F\'{\i}sica, Universidade Federal Fluminense, Av. Litoranea s/n, Gragoat\'{a}, Niter\'{o}i, RJ, 24210-340, Brazil}

\author{J. Lubian} 
	\email{jlubian@id.uff.br}
	\affiliation{Instituto de F\'{\i}sica, Universidade Federal Fluminense, Av. Litoranea s/n, Gragoat\'{a}, Niter\'{o}i, RJ, 24210-340, Brazil}

\author{L.F. Canto}
	\email{canto@if.ufrj.br}
	\affiliation{Instituto de F\'{\i}sica, Universidade Federal do Rio de Janeiro, CP 68528, 21941-972, Rio de Janeiro, RJ, Brazil}

\begin{abstract}
We present a detailed discussion of a recently proposed method to evaluate complete and incomplete fusion cross sections for weakly bound systems. 
The method is applied to collisions of $^{7}$Li projectiles on different heavy targets, and the results are compared with the available data. The overall 
agreement between experiment and theory is fairly good.
\end{abstract}

\maketitle


\section{Introduction}


Fusion reactions involving weakly bound nuclei have attracted considerable interest over the last few decades~\cite{CGD06,KRA07,KAK09,CGD15,KGA16}. The low
breakup threshold of these nuclei influences fusion in two ways. First, the low binding energy of the clusters within the projectile leads to an extended tail in 
the nuclear density, which gives rise to a lower Coulomb barrier. This is a static effect that enhances fusion at all collision energies. Second, the couplings with 
the breakup channel in collisions of these nuclei are very important. They affect elastic scattering and fusion strongly. In addition to the usual direct complete 
fusion (DCF), where the whole projectile fuses with the target, there is incomplete fusion (ICF), where only a piece of the projectile is capture by the target.
Finally, there is the possibility that the projectile breaks up and then all the fragments are absorbed sequentially by the target. This process is known as 
sequential complete fusion (SCF). The sum of DCF and SCF is called complete fusion (CF), and the sum of all fusion processes is called total fusion (TF). \\

The DCF and SCF processes cannot be distinguished experimentally. Besides, most experiments measure only the inclusive TF cross section. However, individual CF and ICF 
cross sections have been measured for some particular projectile-target combinations. There are CF and ICF data available in collisions of $^{6,7}$Li projectiles 
on $^{209}{\rm Bi}$~\cite{DHH02,DGH04}, $^{159}$Tb~\cite{MRP06,BIH75,PMB11}, $^{144,152}$Sm~\cite{RSS13,RSS09,RSS12}, $^{165}$Ho~\cite{TNM02}, 
$^{198}$Pt~\cite{SND13,SNL09}, $^{154}$Sm~\cite{GZH15}, $^{90}$Zr~\cite{KJP12}, $^{124}$Sn~\cite{PSP18}, and $^{197}$Au~\cite{PTN14}.\\

The theoretical determination of individual CF and ICF cross sections has also been a great challenge. Most calculations with this aim are based on classical 
mechanics~\cite{HDH04,DHH02,DHT07,Dia10,Dia11}, or semiclassical approximations~\cite{MCD14,KCD18}, which do not account properly for important
quantum mechanical effects. This shortcoming has been eliminated in a few quantum mechanical models based on the continuum discretized coupled 
channel method (CDCC). However, in most cases, they can only determine the TF cross section~\cite{KKR01,DTB03,JPK14,DDC15}. There is a quantum mechanical method 
that provides individual CF and ICF cross section~\cite{HVD00,DiT02}, but it can only be applied to collisions where the projectile breaks up into
two fragments, with one being much heavier than the other. Recently, Lei and Moro~\cite{LeM19} determined the CF cross sections for the $^{6,7}$Li + $^{209}$Bi systems,
extracting it from the total reaction cross section, by subtracting the inelastic, elastic breakup and inclusive nonelastic breakup (NEB) components. The
NEB cross section was calculated by the spectator-participant model of Ichimura, Austern, and Vincent~\cite{IAV85}. Their method is interesting, but it does not allow 
the calculation of ICF cross sections. There are also the promising quantum mechanical models of Hashimoto {\it et al.}~\cite{HOC09} and of Boseli and 
Diaz-Torres~\cite{BoD14,Bod15}, which in their present stage do not allow quantitative calculations of CF and ICF cross sections.
In a recent letter~\cite{RCL20}, we proposed a new method using CDCC wave functions. This method has the advantage of being applicable to collisions of any
projectile that breaks up into two fragments, independently of their masses.  It was used to evaluate CF and ICF cross sections in the $^7$Li + $^{209}$Bi collision, 
and the results were shown to be in excellent agreement with the data of Dasgupta {\it et al.}~\cite{DHH02,DGH04}.\\

In the present work, we give the details of our method~\cite{RCL20} and use it to evaluate CF and ICF cross sections in collisions of $^{7}$Li with several targets. 
The paper is organized as follows. In section II we introduce our method, expressing the CF, ICF, and TF cross sections
in terms of angular momentum-dependent fusion probabilities, which are calculated in appendix A.  In section III, we present a detailed discussion of the
continuum discretization of $^{7}$Li, in collisions with a $^{209}$Bi target. In section IV, we evaluate CF and ICF cross sections in collisions of $^{7}$Li  projectiles
with $^{209}$Bi, $^{197}$Au, $^{124}$Sn, and $^{198}$Pt targets, and compare the predictions of our method with the experimental data. Finally, in section IV we present our conclusions 
and discuss future extensions of our method.


\section{Theory of Complete and incomplete fusion}\label{theory}


\begin{figure}
\begin{center}
\includegraphics*[width=7 cm]{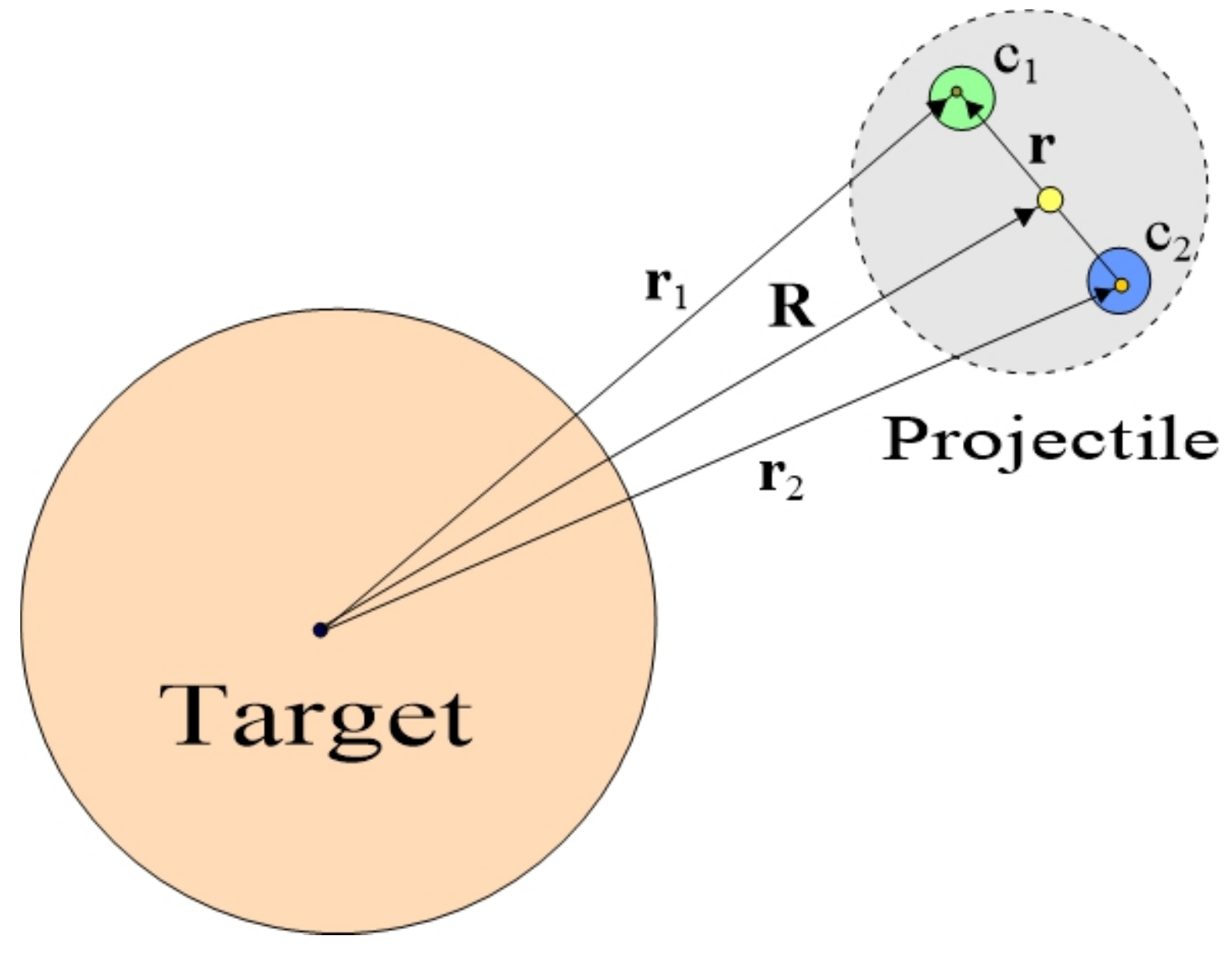}
\end{center}
\caption{(Color on line) Schematic representation of the projectile, its fragments and the target, and the coordinates involved in the calculations.}
\label{F1}
\end{figure}
In this section we describe the theory to evaluate CF and ICF cross sections introduced in Ref.~\cite{RCL20}, which we use in the present work.
We consider the collision of a weakly bound projectile formed by two fragments, $c_1$ and $c_2$, with a spherical target.  The projectile-target
relative vector and the vector between the two fragments of the projectile are denoted by ${\bf R}$ and ${\bf r}$, respectively. For simplicity,
we do not discuss explicitly spins or orbital angular momenta at this stage. The collision dynamics is dictated by the Hamiltonian
\begin{equation}
\mathbb{H}({\bf R},{\bf r}) = h({\bf r}) + \hat{K} + \mathbb{U}^{\scs (1)}(r_{\scs 1})+ \mathbb{U}^{\scs(2)}(r_{\scs 2}) ,
\end{equation}
where 
\begin{equation}
\mathbb{U}^{\scs (i)}(r_{\scr i}) \equiv \mathbb{V}^{\scs (i)}(r_{\scs i}) - i\,\mathbb{W}^{\scs (i)}(r_{\scs i})
\end{equation}
is the complex interaction between fragment $c_{\scs i}$ and the target, with $r_{\scs i}$ representing the distance between their centers. These distances are given by,
\begin{equation}
r_i = \left| {\bf R}+{\bf r}^\prime_i \right|,
\label{distances}
\end{equation}
where ${\bf r}^\prime_i$ is the position vector of fragment $c_i$ in the reference frame of the projectile. For the situation depicted in Fig.~\ref{F1}, these vectors are 
\begin{equation}
 {\bf r}_{\scr 1}^\prime = \frac{A_{\scr 2}}{A_{\scr P}}\, {\bf r}\ \ {\rm and}\ \ 
{\bf r}_{\scr 2}^\prime =  - \frac{A_{\scr 1}}{A_{\scr P}}\, {\bf r},
\label{r1r2}
\end{equation}
with $A_i$ and $A_{\scr P}$ standing for the mass numbers of fragment $c_i$ and the projectile, respectively.

\bigskip

To evaluate the fusion cross sections, we perform CDCC calculations adopting short-range functions for the imaginary potentials $\mathbb{W}^{\scr (1)}$ and $\mathbb{W}^{\scr (2)}$. 
The calculations involve a set of bound channels - subspace B, and a set of continuum-discretized channels ({\it bins}) - subspace C. Since the imaginary potentials have short
ranges, the total fusion cross section is equal to the absorption cross section, which is given by the well known expression~\cite{CaH13}
\begin{equation} 
\sigma_{\scr TF} = \frac{1}{|N|^2}\ \frac{K}{E}\ 
\  \left\langle\ {\rm \Psi}^{\scr (+)}\, \left| \, {\mathbb W}^{(1)} + {\mathbb W}^{(2)}\,  \right| {\rm \Psi}^{\scr (+)}\  \right\rangle.
\label{sigTF-1}
\end{equation}
Above, ${\rm \Psi}^{\scr (+)}$ is the scattering state in a collision with incident wave vector ${\bf K}$ and energy $E$, and $N$ is a normalization
constant. \\

Next, we split the wave function as,\\
\begin{equation} 
{\rm \Psi}^{\scr (+)}({\bf R},{\bf r}) = {\rm \Psi}^{\scr B}({\bf R},{\bf r}) \ +\ {\rm \Psi}^{\scr C}({\bf R},{\bf r}) ,
\end{equation}
where ${\rm \Psi}^{\scr B}$ and ${\rm \Psi}^{\scr C}$ are respectively its components in the bound and bin subspaces. They are given by the channel expansions
\begin{eqnarray} 
{\rm \Psi}^{\scr B}({\bf R},{\bf r}) &=& \sum_{\beta\, \in\, {\scr B}} \ \left[ \psi_{\beta}({\bf R}) \otimes \phi_\beta ({\bf r})\right] \label{PsiB-1}\\
{\rm \Psi}^{\scr C}({\bf R},{\bf r}) &=& \sum_{\gamma\, \in\,  {\scr C}} \ \left[ \psi_{\gamma}({\bf R}) \otimes \phi_\gamma ({\bf r}) \right], 
\label{PsiC-1}
\end{eqnarray}
where $\phi_\beta$ and $\phi_\gamma$ are respectively bound and unbound states of the projectile, and $\psi_\beta$ and $\psi_\gamma$ are
the corresponding wave function describing the projectile-target relative motion.

\bigskip

In our method, we assume that matrix-elements of the imaginary potentials connecting bound channels to bins are negligible. 
Approximations along this line are frequently made in fusion calculations~\cite{SNL87,DTB03,PNT15}. Then, Eq.~(\ref{sigTF-1}) can 
be put in the form
\begin{equation} 
\sigma_{\scr TF} = \sigma_{\scr TF}^{\scr B} \ +\  \sigma_{\scr TF}^{\scr C},
\end{equation}
with
\begin{eqnarray} 
 \sigma_{\scr TF}^{\scr B} &=&  \frac{1}{|N|^2} \frac{K}{E} \sum_{\beta, \beta^\prime\, \in\, {\scr B}} \
 \left\langle \psi_\beta \left|  W_{\beta \beta^\prime}^{\scr (1)}+ W_{\beta \beta^\prime}^{\scr (2)}\,
 \right|   \psi_{\beta^\prime} \right\rangle \label{TF-B}\\
 \sigma_{\scr TF}^{\scr C} &=&  \frac{1}{|N|^2} \frac{K}{E} \sum_{\gamma,\gamma^\prime \in\,{\scr C}} \
 \left\langle \psi_\gamma \left|  W_{\gamma \gamma^\prime}^{\scr (1)} + W_{\gamma \gamma^\prime}^{\scr (2)}\,
 \right|   \psi_{\gamma^\prime} \right\rangle .
 \label{TF-C}
\end{eqnarray}
Above,
\begin{equation} 
W_{\alpha \alpha^\prime}^{\scr (i)} =\left(\phi_{\alpha} \left| \mathbb{W}^{\scr (i)} \right| \phi_{\alpha^\prime} \right),
\label{W form fac}
\end{equation}
with $\alpha, \alpha^\prime$ standing for either $\beta, \beta^\prime$ or $\gamma, \gamma^\prime$, are the matrix-elements of the imaginary potentials.\\

Performing angular momentum expansions of the wave functions and the imaginary potentials, the cross sections of Eqs.~(\ref{TF-B}) and
(\ref{TF-C}) can be put in the form
\begin{eqnarray}
\sigma_{\scr TF}^{\scr B} &=& \frac{\pi}{K^2}\,\sum_J (2J+1)\ \mathcal{P}_{\scr B}^{\scr TF}(J)  \label{TF-B1} \\
\sigma_{\scr TF}^{\scr C} &=& \frac{\pi}{K^2}\,\sum_J (2J+1)\ \mathcal{P}_{\scr C}^{\scr TF}(J) \label{TF-C1},
\end{eqnarray}
with
\begin{eqnarray}
\mathcal{P}_{\scr B}^{\scr TF}(J)  &=& \mathcal{P}_{\scr B}^{\scr (1)}(J) +  \mathcal{P}_{\scr B}^{\scr (2)}(J) \label{P_TF-B} \\
\mathcal{P}_{\scr C}^{\scr TF}(J)  &=& \mathcal{P}_{\scr C}^{\scr (1)}(J) +  \mathcal{P}_{\scr C}^{\scr (2)}(J) \label{P_TF-C} .
\end{eqnarray}
Above, $\mathcal{P}_{\scr B}^{\sc (i)}(J)$ and $\mathcal{P}_{\scr C}^{\sc (i)}(J)$ are the probabilities of absorption of fragment $c_i$ in
bound channels and in the continuum, respectively. They are the contributions of $\mathbb{W}^{\scr(i)}$ to the TF cross section.
A detailed calculation of these quantities is presented in the appendix. \\

Since $\sigma_{\scr TF}^{\scr B}$ is a sum of contributions from bound channels, we assume that the two fragments are absorbed simultaneously.
Thus, we write
\begin{equation}
\sigma_{\scr DCF} = \sigma_{\scr TF}^{\scr B} .
\label{DCF}
\end{equation}

\bigskip

The meaning of $\sigma_{\scr TF}^{\scr C}$ is not so clear. Since it is a sum of contributions from unbound channels, it must be 
related to cross sections of the ICF and SCF processes. Thus, the individual cross sections for ICF of fragment $c_{\scr i}$ (ICFi) and for SCF can be written as
\begin{equation}
\sigma_{\scr ICFi} = \frac{\pi}{K^2}\,\sum_J (2J+1)\ \mathcal{P}^{\scr ICFi}(J) 
\label{sigICFi}
\end{equation}
and 
\begin{equation}
\sigma_{\scr SCF} = \frac{\pi}{K^2}\,\sum_J (2J+1)\ \mathcal{P}^{\scr SCF}(J) ,
\label{SCF}
\end{equation}
where the ICF probabilities, $\mathcal{P}^{\scr ICFi}(J)$, and the SCF probability, $\mathcal{P}^{\scr SCF}(J)$, are functions of the absorption 
probabilities $\mathcal{P}_{\scr C}^{\scr (1)}(J)$ and $\mathcal{P}_{\scr C}^{\scr (2)}(J)$. These functions will be determined in the next sub-section.\\

The CF, ICF and TF cross sections are then given by
\begin{eqnarray}
\sigma_{\scr CF} &=& \sigma_{\scr DCF}\,+\,\sigma_{\scr SCF},  \label{sigCF} \\
\sigma_{\scr ICF} &=&\sigma_{\scr ICF1} \,+\, \sigma_{\scr ICF2}, \label{sigICF}\\
\sigma_{\scr TF} &=& \sigma_{\scr CF} + \sigma_{\scr ICF} . \label{sigmaTF-sum} 
\end{eqnarray}
%


\subsection{$J$-dependent elastic, nonelastic and absorption probabilities}


We consider a coupled channel problem involving the elastic channel ($\alpha = 0$) and $N$ nonelastic channels ($\alpha = 1,2,...,N$). 
The absorption cross section is given in terms of the total reaction cross section and the cross sections for non-elastic channels
by the equation
\begin{equation}
\sigma_{\rm abs}  =   \sigma_{\scr R} - \sum_{\alpha = 1}^N \sigma_\alpha.
 \label{abs 1}
\end{equation}
Carrying out angular momentum expansions, we get 
\begin{equation}
\sigma_{\rm abs}  =  \sum_{J=0}^ \infty \sigma_{\rm abs}(J)  =  \sum_{J=0}^ \infty \left[
\sigma_{\scr R} (J)- \sum_{\alpha = 1}^N \sigma_\alpha(J)
\right],
 \label{abs 2}
\end{equation}
with 
\begin{eqnarray}
\sigma_{\scr R} (J)  &=& \frac{\pi}{K^2}\ (2J+1)\ \Big[ 1- \left| S_{0}(J) \right|^2 \Big]\label{sigR-1}\\
\sigma_{\alpha} (J)  & =& \frac{\pi}{K^2}\   (2J+1)\ \left| S_{\alpha}(J) \right|^2.
\label{sigalpha-1}
\end{eqnarray}
Then, the $J$-components of the absorption cross section are given by,
\begin{equation}
\sigma_{\rm abs}(J)  = A(J)\ \Bigg[ 1\ -\  \sum_{\alpha=0}^N  \left| S_{\alpha}(J) \right|^2 \Bigg]
 \label{abs 3}
\end{equation}
where
\begin{equation}
 A(J) =  2\pi\ \left( \frac{\Lambda}{K} \right)\, \left( \frac{1}{K} \right).
 \label{AJ}
\end{equation}
Above, we have introduced the semiclassical angular momentum in $\hbar$ units, $\Lambda = J+1/2$. 
The two terms within brackets in Eq.~(\ref{AJ}) correspond respectively to the impact parameter, $b$, and its increment, $\Delta b$, when $\Lambda$ is 
increased by one unit. Thus, $A(J)$ is the area of a ring with radius $b$ and thickness $\Delta b$. Therefore, $\mathcal{P}_{\alpha}(J)\equiv \left| S_\alpha(J) \right|^2$
is the probability that the system is  in channel-$\alpha$ after a collision with angular momentum $J$. Then, Eq.~(\ref{abs 3}) leads to the relation,
\begin{equation}
\mathcal{P}_{\rm abs}(J) \equiv \frac{\sigma_{\rm abs}(J)}{A(J)} = 1\,-\,  \sum_{\alpha=0}^N  \mathcal{P}_\alpha(J),
 \label{abs 4}
\end{equation}
and one gets the normalization relation
\begin{equation}
\mathcal{P}_{\rm abs}(J) +  \sum_{\alpha=0}^N  \mathcal{P}_\alpha(J) = 1.
 \label{normaliz-0}
\end{equation}
%

\subsubsection{Probabities in the CDCC calculation}


In our CDCC calculations the target is treated as a heavy inert particle. Then, the $N+1$ channels in the sum of Eq.~(\ref{normaliz-0}) differ by the state of the projectile.
The first term is the elastic channel ($\alpha=0$). The remaining $N$ channels can be split as $N=N_{\scr B} + N_{\scr C}$, where $N_{\scr B}$ is the number of inelastic channels 
and $N_{\scr C}$ is the number of bins in the continuum discretization. Then, the sum over the excited states gives the total inelastic probability and the sum over the bin states 
the elastic breakup probability. That is,
\begin{equation}
\mathcal{P}^{\rm el}(J) = \mathcal{P}_{0}(J),\ \ \ \mathcal{P}^{\rm inel}(J) = \sum_{\alpha=1}^{N_{\scr B}}\  \mathcal{P}_{\alpha}(J)
 \label{el-inel}
\end{equation}
and 
\begin{equation}
 \mathcal{P}^{\scr EBU}(J) =  \sum_{N_{\scr B}+1}^{N}\ \mathcal{P}_{\alpha}(J).
 \label{Pbu}
\end{equation}

Since the imaginary potentials in our calculations have short range, absorption represents fusion, of any kind, namely $\mathcal{P}_{\rm abs}(J) = \mathcal{P}^{\scr TF}(J)$.
The normalization condition of Eq.~(\ref{normaliz-0}) then reads,
\begin{equation}
\mathcal{P}^{\scr TF}(J)\,+\, \mathcal{P}^{\rm el}(J)\, +\, \mathcal{P}^{\rm inel}(J)\, +\,  \mathcal{P}^{\scr EBU}(J) = 1.
 \label{normaliz-1}
\end{equation}
The probabilities $\mathcal{P}^{\rm el}$, $\mathcal{P}^{\rm inel}$ and $\mathcal{P}^{\scr EBU}$ are directly given by the solution of the CDCC equations. The TF probabilites
are evaluated by the angular momentum projected version of Eq.~(\ref{sigTF-1}) (see appendix A).\\

\subsubsection{ICF and SCF Probabities}

The contribution from the continuum to the TF probability is 
\begin{equation}
\mathcal{P}^{\scr TF}_{\scr C}(J) = \mathcal{P}^{\scr ICF1}(J) \,+\, \mathcal{P}^{\scr ICF2}(J)\,+\, \mathcal{P}^{\scr SCF}(J).
 \label{TF-ICF1-ICF2-SCF}
\end{equation}
However, to evaluate the above probabilities, they must be expresses in terms of absorption probabilities of the two fragments, $\mathcal{P}_{\scr C}^{\scr (1)}(J)$
and $\mathcal{P}_{\scr C}^{\scr (2)}(J)$, which are calculated in appendix A. Following Refs.~\cite{MCD14,KCD18,RCL20}, we make the intuitive assumptions
\begin{eqnarray}
 {\mathcal P}^{\scr ICF1}(J)  &=& {\mathcal P}^{\scr (1)}_{\scr C} (J) \times \left[\ 1 - {\mathcal P}^{\scr (2)}_{\scr C} (J)\ \right]   \label{PICF1}\\
{\mathcal P}^{\scr ICF2}(J)  &=&  {\mathcal P}^{\scr (2)}_{\scr C} (J) \times \left[\ 1 - {\mathcal P}^{\scr (1)}_{\scr C} (J)\ \right]   \label{PICF2}.
\end{eqnarray}
The SCF probability is then obtained inserting Eqs.~(\ref{P_TF-C}), (\ref{PICF1}) and (\ref{PICF2}) into Eq.~(\ref{TF-ICF1-ICF2-SCF}). We get
\begin{equation}
{\mathcal P}^{\scr SCF}(J)  =  2\  {\mathcal P}^{\scr (1)}_{\scr C} (J) \times  {\mathcal P}^{\scr (2)}_{\scr C} (J).
 \label{TSCF}
\end{equation}
Note that the factor 2 is essential to satisfy Eq.~(\ref{P_TF-C}). In fact, it should be expected since differences in the order of events in the sequential absorption of 
the two fragments must involve different intermediate states.


\section{Applications}


We used our method to study fusion reactions in collisions of  $^{7}{\rm Li}$ projectiles with $^{209}$Bi, $^{197}$Au, $^{124}$Sn,
and $^{198}$Pt  targets, for which experimental data are available. In our calculations,  $^{7}{\rm Li}$ is treated as the two-cluster system:
$^7{\rm Li}\, \equiv\, ^3{\rm H} +\, ^4{\rm He}$, with separation energy $B = 2.45$ MeV. To determine the cross sections, we used the CF-ICF 
computer code (unpublished), which evaluates the angular momentum projected version of the expressions of the previous section, derived
in the appendix. These expressions involve intrinsic 
states of the projectile and radial wave functions, which were obtained running the CDCC version of the FRESCO code~\cite{Tho88}.\\

The real part of the interaction between fragment $c_i$ and the target, $\mathbb{V}^{\sc (i)}(r_i)$, is given by the S\~ao Paulo potential~\cite{CPH97}
(SPP), calculated with the densities of the systematic study of Chamon {\it et al.}~\cite{CCG02}. The projectile-target potential in the elastic channel
is then given by
\begin{equation}
V_{\scr 00}(R) = \int d^3{\bf r}\, \left| \phi_0({\bf r})  \right|^2\ 
\left[  
\mathbb{V}^{\sc (1)}(r_1) \,+\, \mathbb{V}^{\sc (2)}(r_2) \right] ,
\label{V00}
\end{equation}
where $ \phi_0({\bf r})$ is the ground state wave function of the projectile. Note that this potential takes into account the low breakup threshold 
of the projectile. This makes its Coulomb barrier lower than the one given by the SPP calculated directly for the projectile-target system. This static 
effect of the low binding energy enhances the fusion cross section below and above the barrier.\\

Since the imaginary part of the fragment-target potentials represent fusion absorption, they must be strong and act exclusively in the inner region 
of the Coulomb barrier. Then, we adopted Woods-Saxon functions with the form,\\
\begin{equation}
\mathbb{W}^{(i)}(r_i) =  \frac{W_0}{1+\exp\left[\left( r_i - R_{\rm w} \right) /  a_{\scr w}\right]},\qquad i=1,2,
\label{Wi}
\end{equation}
with the following parameters  
\begin{equation}
W_0 = 50\, {\rm MeV}, \ \ \  R_{\rm w} = 1.0\,\left[  A_i^{\scr 1/3} +  A_{\scr T}^{\scr 1/3} \right]\,{\rm  fm};\, \  a_{\scr w} = 0.2\,{\rm fm}.
\label{par-Wi}
\end{equation}

The intrinsic states of the projectile are solutions of a Schr\"odinger equation with the Hamiltonian
\begin{equation}
h({\bf r}) = K_{\bf r}+V_{\scr 12}(r_{\scr 12}),
\label{hr}
\end{equation}
where $K_{\bf r}$ is the relative kinetic energy of fragments within the projectile, and $V_{\scr 12}(r_{\scr 12})$ is the interaction potential between them. The 
potential used to describe the bound states of the projectile was parametrized by Woods-Saxon functions and derivatives (for the spin-orbit term), with
parameters fitted to reproduce its binding energy. Different potentials were used for continuum states. In this case, the parameters
were fitted to reproduce the energies and widths of the main resonances. The parameters are basically the ones adopted by Diaz-Torres, 
Thompson and Beck~\cite{DTB03}, except for the reduced radius of the central potential. We used $ r_{\scr 0} =1.153$ fm, that gives a slightly better
description of the resonances of $^ 7$Li. Their experimental energies and widths are shown in Table \ref{res-7Li}, together with the theoretical values 
obtained in this way.\\
\begin{table}
\centering
\caption{ Experimental~\cite{nndc} and theoretical energies and widths of the  $^7$Li resonances. The energies and widths are given
in MeV. }
\vspace{0.2cm}
\begin{tabular}{cccccc}
\hline 
\ \ \ \ \ \ $l$\qquad\qquad &\;\;$j^\pi$\ \ \;\; &\;\; $\varepsilon^{\scr th}_{\scr res}$ \;\; & \;\;$\ \Delta_{\rm th}$ \;\; &\ \;\; $\varepsilon_{\rm res}^{\scr exp}$\;\; 
&\  \;\; $\Delta_{\scr exp}$\;\; \\ 
\hline                               
3 & $7/2^{-}$ &2.15  & 0.1    & 2.16   &0.093 \\
\hline
3 & $5/2^{-}$ & 4.54 & 0.88  & 4.21  &0.88 \\
\hline
\end{tabular}
\label{res-7Li}
\end{table}

Multipole expansions of the potentials were carried out, taking into account multipoles up to $\lambda = 4$. In the CDCC calculations we used a
matching radius of 40 fm and considered total angular momenta up to $J = 60\,\hbar$. Note that higher angular momenta, which are essential in
calculations of breakup cross sections, do not give relevant contribution to fusion. We checked the convergence of the calculations with respect 
to these parameters and found that the results are very stable.\\


\subsection{Discretization of the continuum}

%

The channel expansion of Eq.~(\ref{PsiB-1}) included the ground state of $^7$Li ($j=3/2^-, l=1$) and its only excited state, with energy 
$\varepsilon^* = 0.48$ MeV ($j=1/2^-, l=1$). 

\medskip

The continuum expansion of Eq.~(\ref{PsiC-1}) included bins generated by scattering states of the $^3{\rm H}\,-\,^4{\rm He}$ system, with 
orbital angular momenta $l=0,...,l_{\rm max}$ ($1/2 \le j \le l_{\rm max}+1/2$) and collision energies from zero to a cut-off energy $\varepsilon_{\rm max}$.
The bins were generated by the equation
\begin{equation}
u_{\scs \beta\,  l_\beta j_\beta}(r) = \int d\varepsilon\ \Gamma_{\scr \beta}(\varepsilon)\, u_{\scs \varepsilon l_\beta j_\beta}(r),
\label{functions phi}
\end{equation}
where $u_{\varepsilon l_\beta j_\beta}(r)$ is the radial wave function in a scattering state with collision energy $\varepsilon$, and angular momentum quantum
numbers $l_\beta ,j_\beta$, and $\Gamma_{\scr \beta}(\varepsilon)$ is a weight function concentrated around the energy $\varepsilon_{\scr \beta}$. In the present 
work we discretize the continuum in the energy space, using bins with constant values within some interval around $\varepsilon_\beta$. Weight functions 
of this kind, either in the energy or in the momentum space, are commonly used in the literature~\cite{SYK86,AIK87,MKO03,ThN09}. The weight functions
were given by
\begin{eqnarray}
 \Gamma_{\scr \beta}(\varepsilon) & = & \frac{1}{\sqrt{\Delta_{\scr \beta}}}, \qquad {\rm if}\ \varepsilon_{\scr \beta}^{\scr (+)} \ge 
 \varepsilon_{\scr \beta} \ge \varepsilon_{\scr \beta}^{\scr (-)} 
\nonumber \\
                                         & = & 0, \qquad\qquad\   {\rm otherwise}.
\label{constant gamma}
\end{eqnarray}
\begin{figure}
\begin{center}
\includegraphics*[width=7cm]{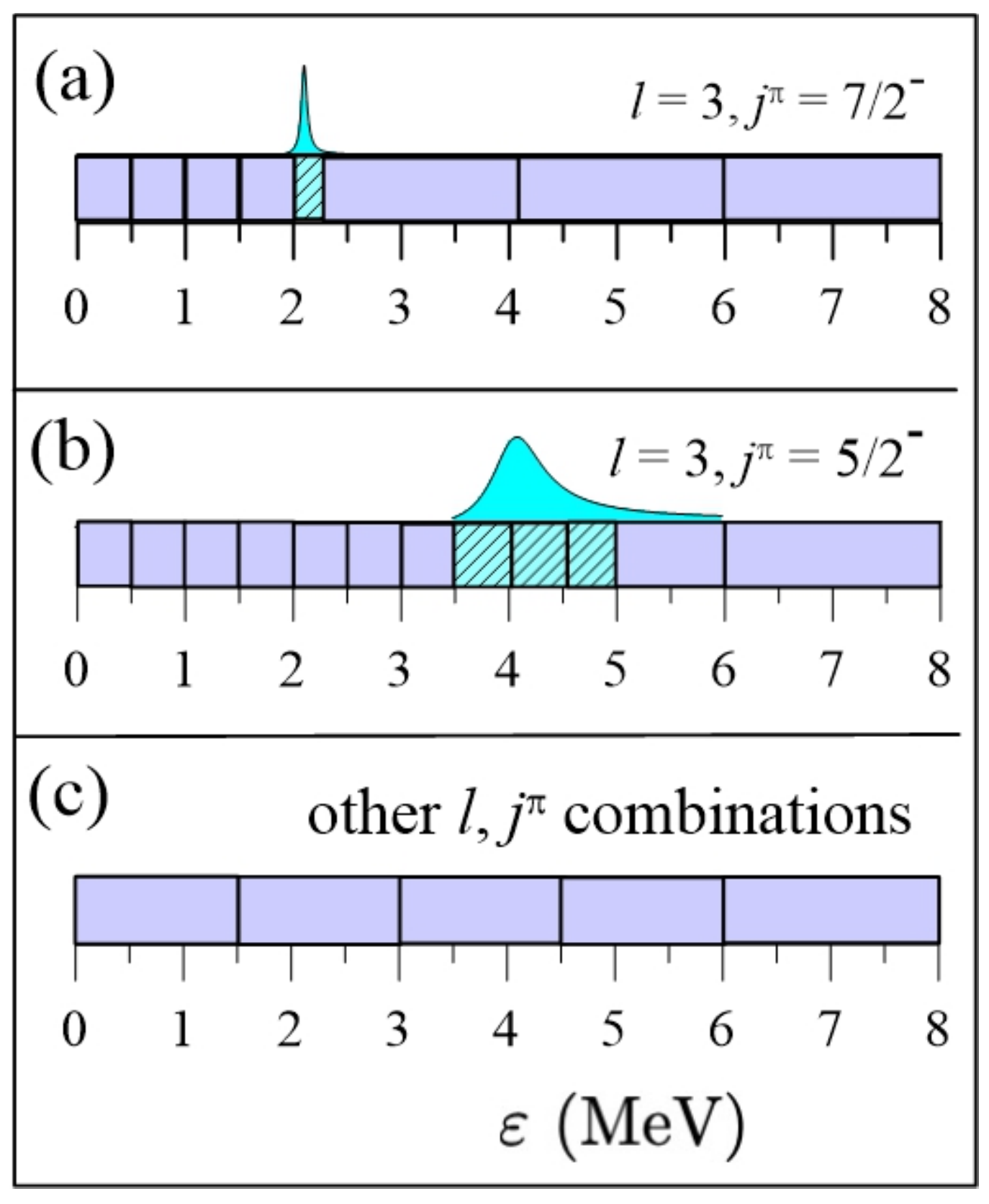}
\end{center}
\caption{(Color on line) Discretization of the continuum of $^7$Li (panel (a)) and $^7$Li (panel (b)). The narrower bins in the resonances regions 
are represented in light blue.}
\label{F2}
\end{figure}
Above, $\varepsilon_{\scr \beta}^{\scr (\pm)} = \varepsilon_{\scr \beta}\, \pm\, \Delta_{\scr \beta}/2$ are the limits of the interval. The bins must cover the
whole energy interval from zero to $\varepsilon_{\rm max}$. That is, the upper limit of the $\beta^{\rm th}$ bin, 
$\varepsilon_{\scr \beta}\, + \, \Delta_{\scr \beta}/2$, should coincide with the lower limit of the subsequent bin, 
$\varepsilon_{\scr \beta + 1}\, -\, \Delta_{\scr \beta +1}/2$. \\

The locations and widths of the bins depend on the resonance structure of the projectile. In the absence of resonances, good convergence can be achieved 
using bins with $\Delta \sim 1 - 2$ MeV, or even larger than this. To increase the speed of the numerical calculations, the number of bins can be reduced 
using broader bins  as $\varepsilon$ approaches $\varepsilon_{\rm max}$. The situation is more complicated in the presence of sharp resonances. Then, 
it is necessary to use at least one narrow bin in the resonance region. The meshes for angular momenta with and without resonances are represented in 
Fig.~\ref {F2}. For $l=3,j^\pi = 7/2^{\scr -}$, where there is a sharp resonance at $\varepsilon_{\scr res} = 2.16$ MeV, with $\Delta_{\rm exp} =  0.093$ MeV (see  
Table \ref{res-7Li}), we used the mesh represented in panel (a). The region below the resonance comprised 4 bins of $\sim 0.5$ MeV, and the resonance was
covered by a single bin of width 0.2 MeV. Above the resonance, we used 3 bins of width $\sim 2$ MeV. For $l=3,j^\pi = 5/2^{\scr -}$ there is a broader 
resonance at $\varepsilon_{\rm res} = 4.21$ MeV, with $\Delta_{\rm exp} =  0.88$ MeV (see Table \ref{res-7Li}). 
Then, we adopted the mesh represented in panel (b). Below the resonance, we used 7 bins with $\sim \Delta = 0.5$ MeV. The resonance region, between 
3.5 and 5 MeV, was covered by 3 bins of about the same width, and the region between 5 and 8 MeV was covered by a bin of 1 MeV and a bin of 2 MeV. 
Finally in the remaining cases, where there are no resonances, the continuum was discretized with 4 bins of  $\Delta = 1.5$ MeV and one bin of 
$\Delta = 2.0$ MeV, as shown in panel (c).\\

\begin{figure}
\begin{center}
\includegraphics*[width=6.5cm]{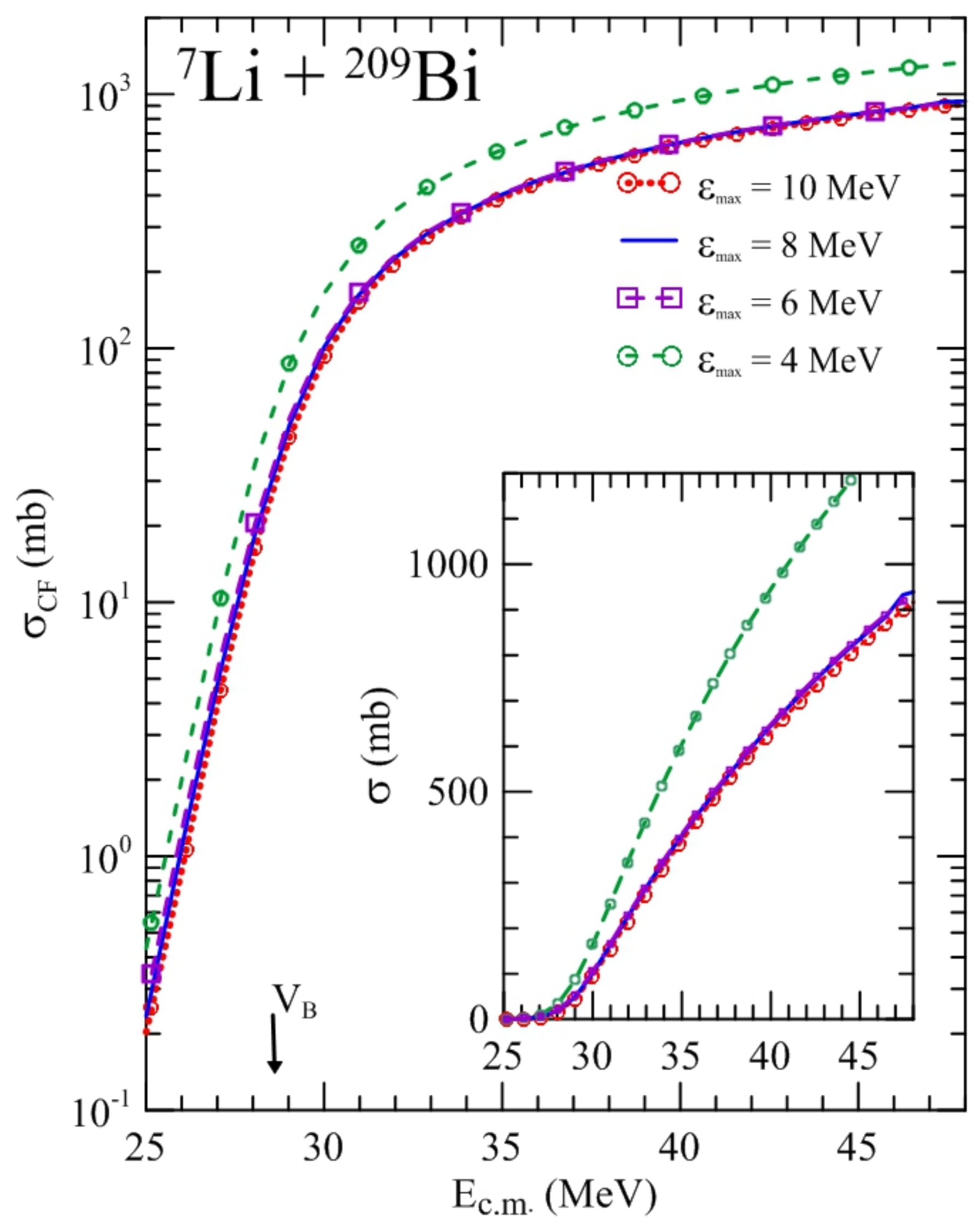}
\end{center}
\caption{ (Color on line) Convergence of $\sigma_{\scr CF}$ with respect to $\varepsilon_{\rm max}$.}
\label{F3}
\end{figure}
\begin{figure}
\begin{center}
\includegraphics*[width=6.5cm]{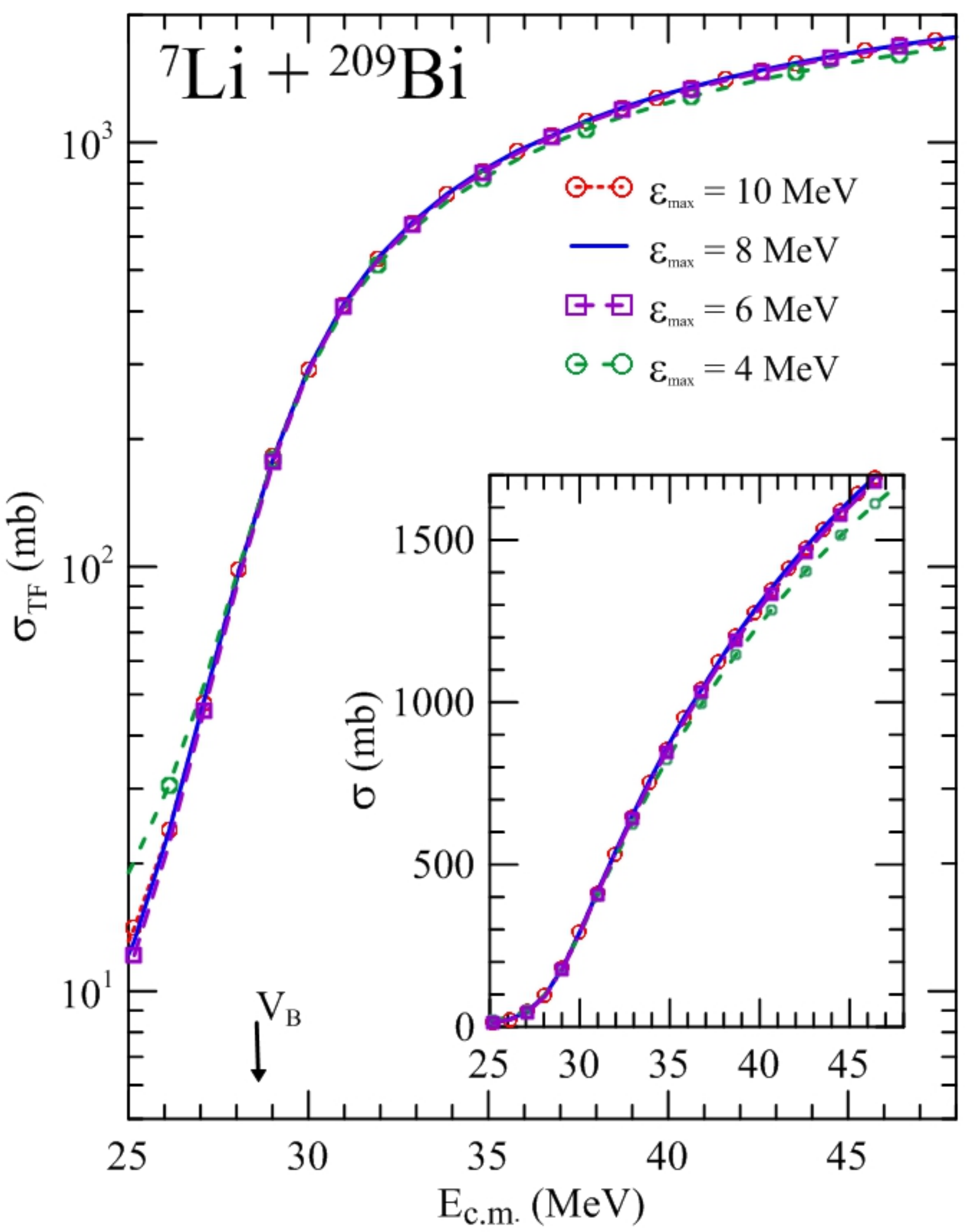}
\end{center}
\caption{(Color on line) Convergence of $\sigma_{\scr TF}$ with respect to $\varepsilon_{\scr max}$.}
\label{F4}
\end{figure}
\begin{figure}
\begin{center}
\includegraphics*[width=6.5cm]{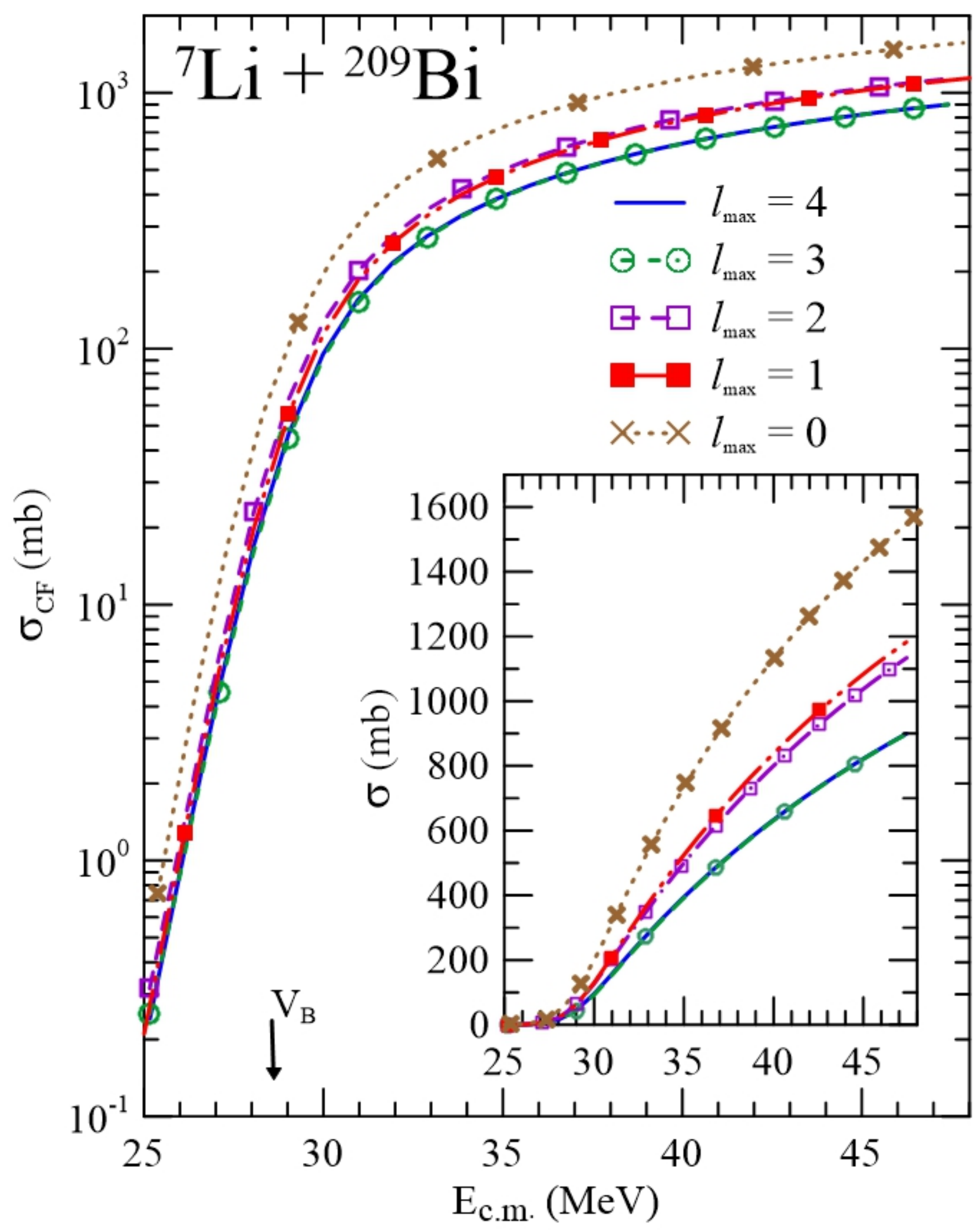}
\end{center}
\caption{(Color on line) Convergence of $\sigma_{\scr CF}$ with respect to $l_{\scr max}$.}
\label{F5}
\end{figure}
\begin{figure}
\begin{center}
\includegraphics*[width=6.5cm]{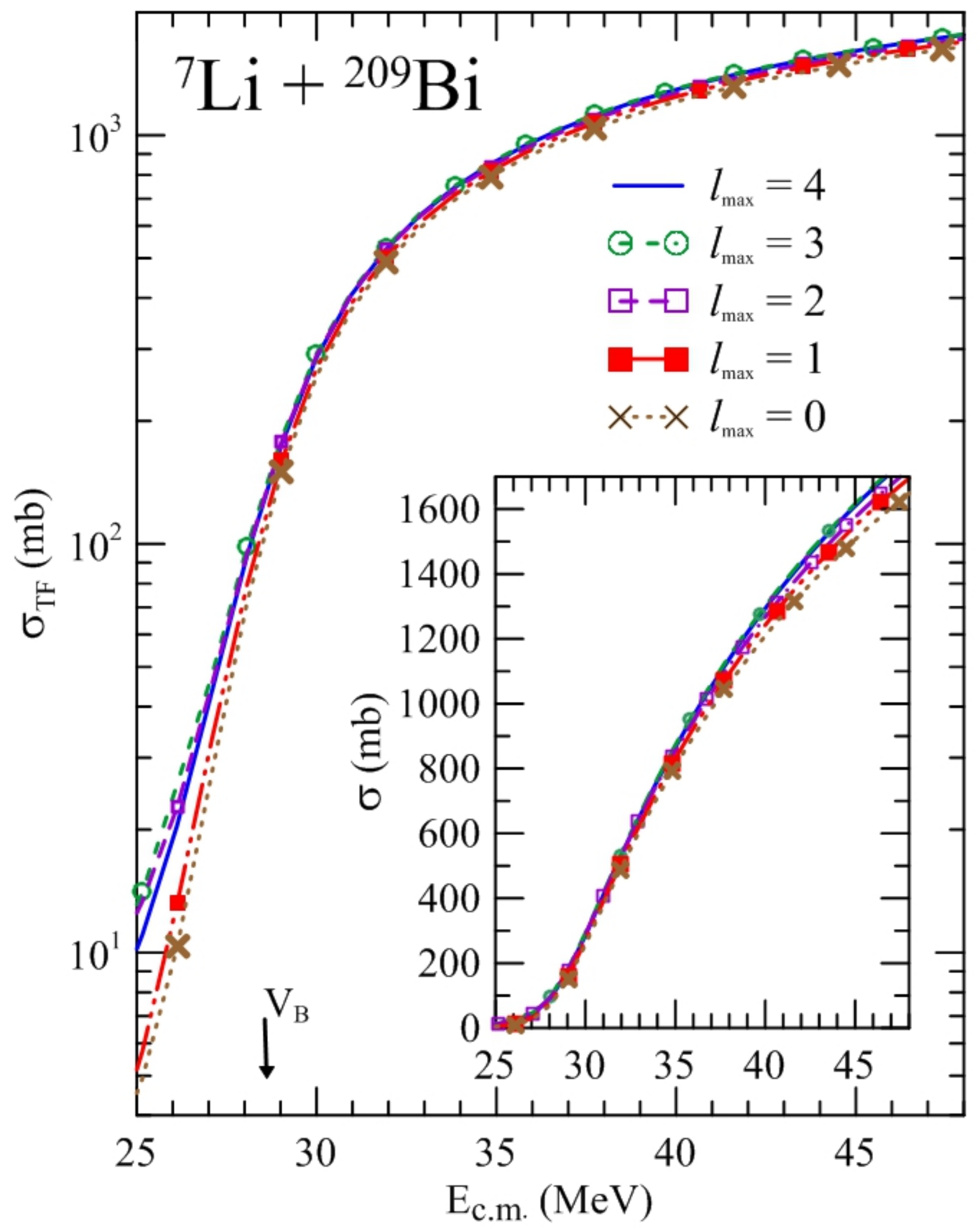}
\end{center}
\caption{(Color on line) Convergence of $\sigma_{\scr TF}$ with respect to $l_{\scr max}$.}
\label{F6}
\end{figure}
We got very good convergence in our calculations using $\varepsilon_{\rm max} = 8$ MeV and $l_{\rm max} = 3\,\hbar$. This is illustrated in Figs.~\ref{F3}
to \ref{F6}, which show cross sections of the $^7{\rm Li}\,+\,^{209}{\rm Bi}$ system, for different values of $\varepsilon_{\rm max}$ and $l_{\rm max}$. 
The main body of the figures shows cross sections in logarithmic scales, whereas the insets show results in linear scales. In this way, the convergence 
below and above the barrier can be easily assessed. Inspecting Fig.~\ref{F3}, one concludes that the convergence of $\sigma_{\scr CF}$ for 
$\varepsilon_{\rm max} = 8$ MeV is excellent. The cross section can hardly be distinguished from the one obtained with
the higher cut-off value of $\varepsilon_{\rm max} = 10$ MeV. Even for $\varepsilon_{\scr max} = 6$ MeV, the convergence is already quite good. The situation 
for $\sigma_{\scr TF}$, shown in Fig. \ref{F4}, is similar, with the convergence above the barrier being still better.
The convergence of the CF and TF cross sections with respect to $l_{\rm max}$, illustrated respectively in Figs. \ref{F5} and \ref{F6}, is also very good. 
In both cases, the results obtained with $l_{\rm max} = 3\,\hbar$ can hardly be distinguished from those obtained with $l_{\rm max} = 4\,\hbar$. \\

Although the above discussion has been restricted to the $^7{\rm Li}\,+\,^{209}{\rm Bi}$ system, similar behaviors were found for the other system considered
in the present work. In all cases we got good convergence with the same discretization of the continuum. \\

We remark that the convergence study presented above involves the usual parameters of the CDCC method, $\varepsilon_{\rm max}$ and $l_{\rm max}$, 
which define the truncation of the continuum space. There are, however, internal parameters of FRESCO, related to numerical procedures adopted within 
the code. Typical applications of FRESCO are calculations of direct reaction cross sections, which depend exclusively on the components of
the S-matrix, given by the asymptotic form of the radial wave functions. In such cases, it is not necessary to change the default values of the internal
parameters of the code. The situation is more complex in the present work. As shown in appendix A, the CF and ICF cross sections of our method are
expressed in terms of radial integrals of the short-range imaginary potentials, multiplied by radial wave functions. Since the main contributions to these 
integrals come from small radial distances, the asymptotic convergence of the radial wave functions is not enough. One has to make sure that the radial 
wave functions are stable in the inner region of the barrier, where they are very small. For this purpose, it may be necessary to modify the default value
of these parameters.


\subsection{Spectroscopic amplitudes}\label{sect Spec-Fac}


In the calculation of matrix-elements between bound channels and continuum-discretized states, the latter have the
$^3{\rm H}\,-\,^4{\rm He}$ cluster configuration intrinsically, and so does the interaction $\mathbb{V}^{\scr (1)} + 
\mathbb{V}^{\scr (2)}$. However, the bound states of $^7$Li do not. Although the amplitude for this configuration is expected to be
dominant, it is definitely not equal to one. This statement is supported by the large cross sections for transfer
reactions of a single nucleon, observed in collisions of this nucleus~\cite{RRL10,LDH11,LDH13,ZZH18}. The probabilities of
finding the dominant cluster configuration in $^{6,7}$Li is expected to be of the order of 70\%~\cite{WMO15}. Then, 
the bound-continuum matrix elements should be multiplied by some spectroscopic amplitude, $\mathcal{S}$, say in the 
$\{0.7 - 1.0\}$ range. This amplitude could be neglected in qualitative calculations, but not if one aims at a quantitative
description of the data. \\

\begin{figure}[t]
\begin{center}
\includegraphics*[width=7 cm]{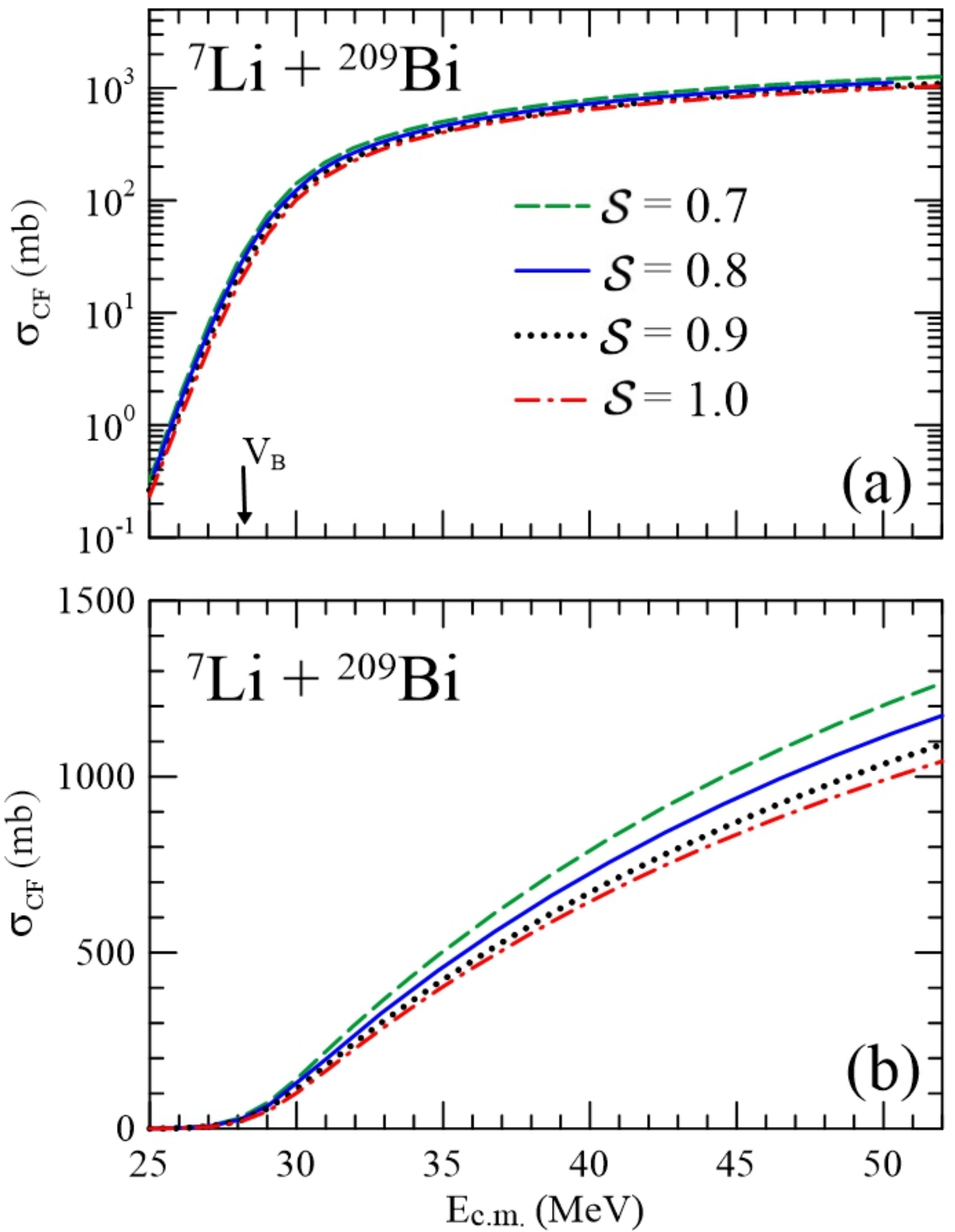}
\end{center}
\caption{(Color on line) CF cross sections calculated with different values of the spectroscopic amplitude.}
\label{F7}
\end{figure}
Since the inclusion of the spectroscopic amplitude weakens the couplings with the breakup channel, it is expected to enhance the
DCF cross section and suppress ICF. The former effect is illustrated in Fig.~\ref{F7}, that shows CF cross sections calculated with 
the spectroscopic  amplitudes: $\mathcal{S} = 0.7, 0.8, 0.9$ and 1.0. The results are show in logarithmic (panel (a)) and linear scales (panel (b)).
In the logarithmic plot, the curves for the different spectroscopic amplitudes can hardly be distinguished. However, the influence of 
$\mathcal{S}$ can be observed in the linear plot. For variations of  $\mathcal{S}$ in the $\{ 0.7,1.0 \}$ range, the cross section changes up to 
$\sim 20\%$. Unfortunately, there are no accurate calculations of the spectroscopic amplitude. Then, we treat it as a free parameter, that 
can vary between 0.7 and 1.0. \\

Deviations of the bound states of the projectile from the $^3{\rm H}\,-\,^4{\rm He}$ cluster configuration may also affect diagonal matrix
elements of the interaction. They are expected to modify the barrier of the $V_{\scr 00}(R)$ potential. However, such effects 
are not expected to be very important. This potential is basically determined by the densities of the collision partners, and it is very sensitive to the long tail of
the projectile's density. This has been taken into account, through the use of a $V_{12}$ potential that reproduces the experimental binding energy of $^7$Li. 
Although a more careful study of this problem is called for, we will leave it to a future work.\\


\subsection{Complete fusion cross sections}\label{sect CF}

%
\begin{figure}[t]
\begin{center}
\includegraphics*[width=7 cm]{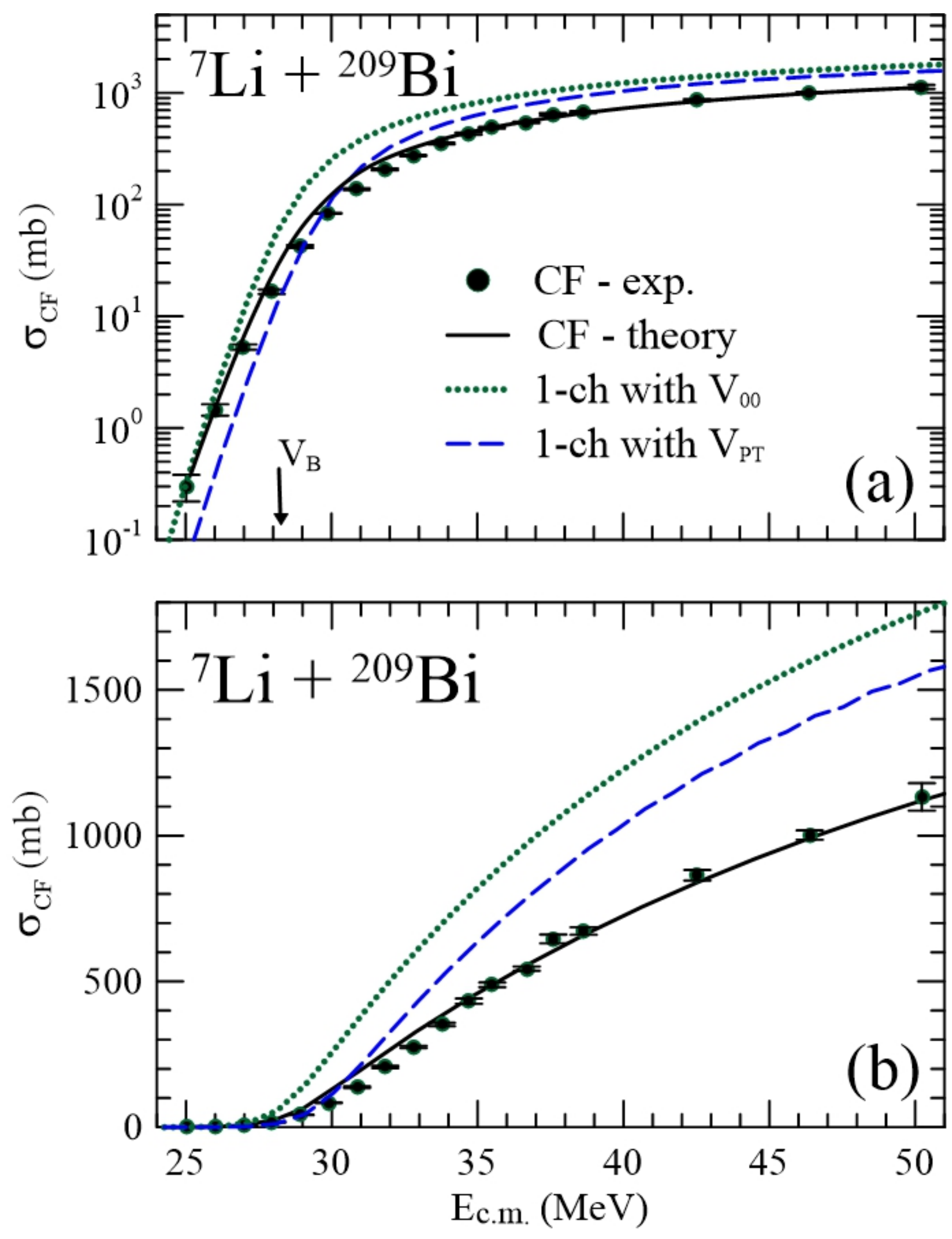}
\end{center}
\caption{(Color on line) Calculated CF cross sections for the $^7$Li + $^{209}$Bi system (solid black line) in comparison with the data of 
Refs.~\cite{DHH02,DGH04} (open circles). }
\label{F8}
\end{figure}
\begin{figure}
\begin{center}
\includegraphics*[width=7 cm]{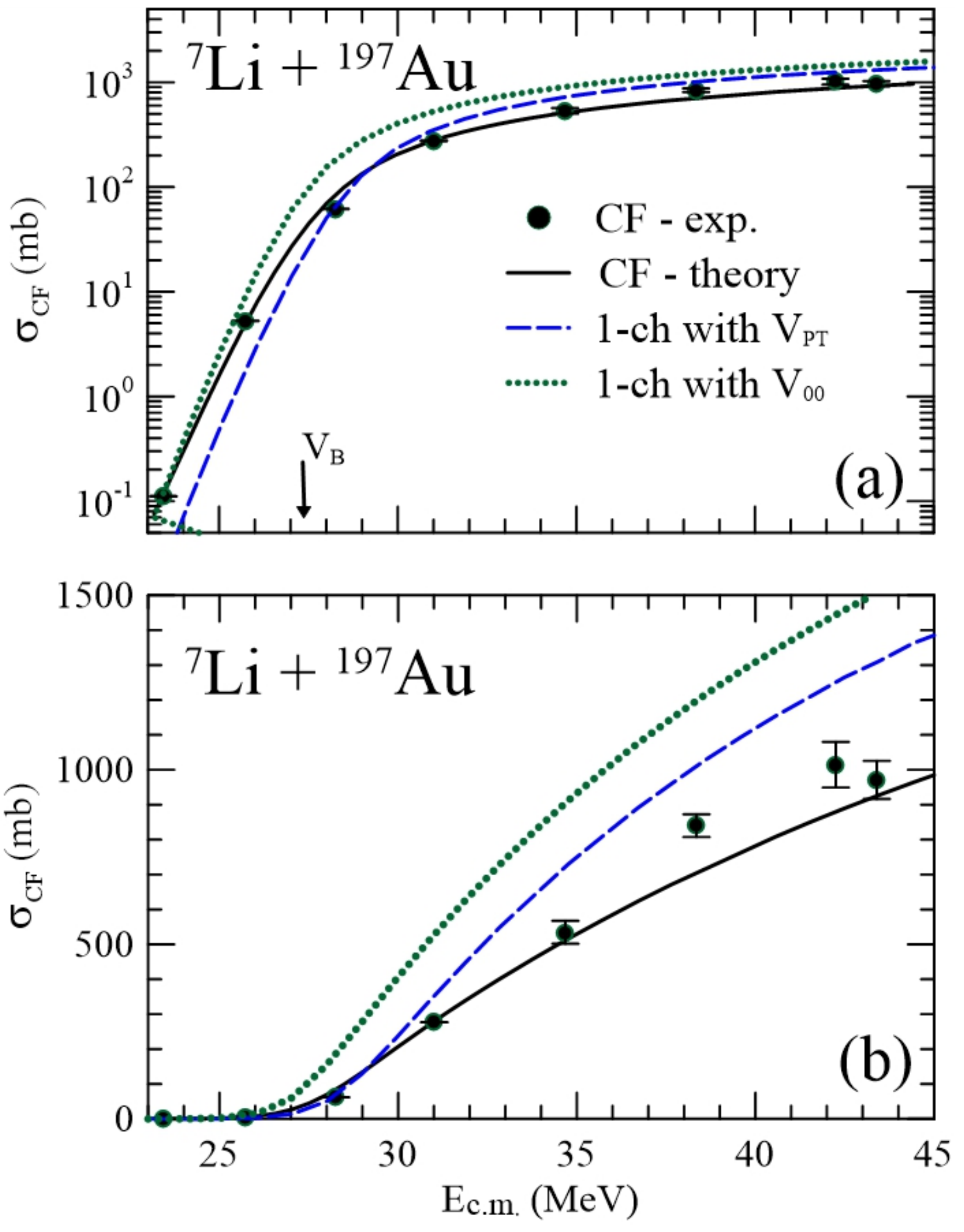}
\end{center}
\caption{(Color on line) Same as Fig.~\ref{F8}, but for the $^7$Li + $^{197}$Au system. Here the data are from Refs.~\cite{PTN14,nndc}.}
\label{F9}
\end{figure}
\begin{figure}
\begin{center}
\includegraphics*[width=7 cm]{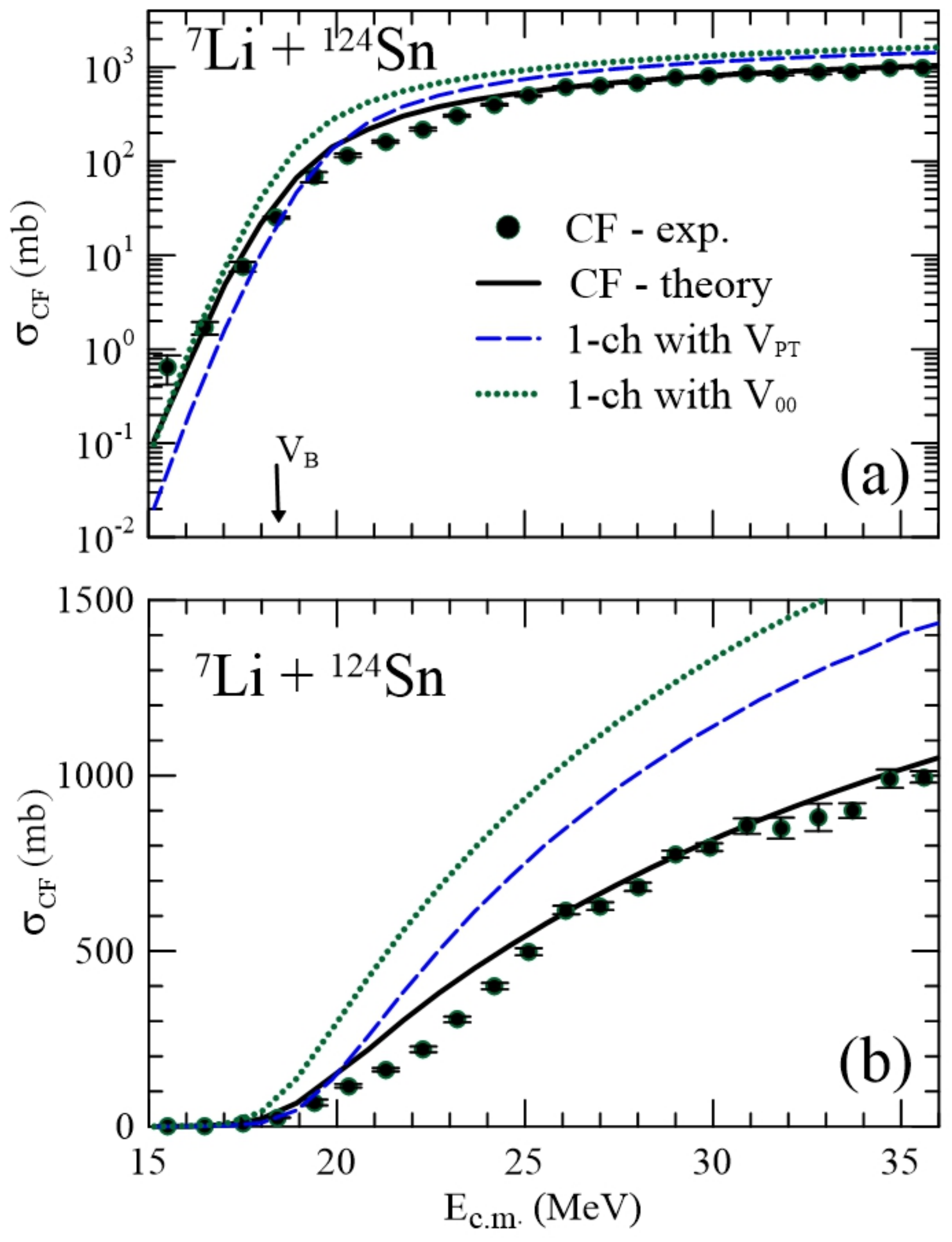}
\end{center}
\caption{(Color on line) Same as Fig.~\ref{F8} but now the system is  $^7$Li + $^{124}$Sn. Here, the system is $^7$Li + $^{124}$Sn and the data are from
Ref.~\cite{PSP18}.}
\label{F10}
\end{figure}
\begin{figure}
\begin{center}
\includegraphics*[width=7 cm]{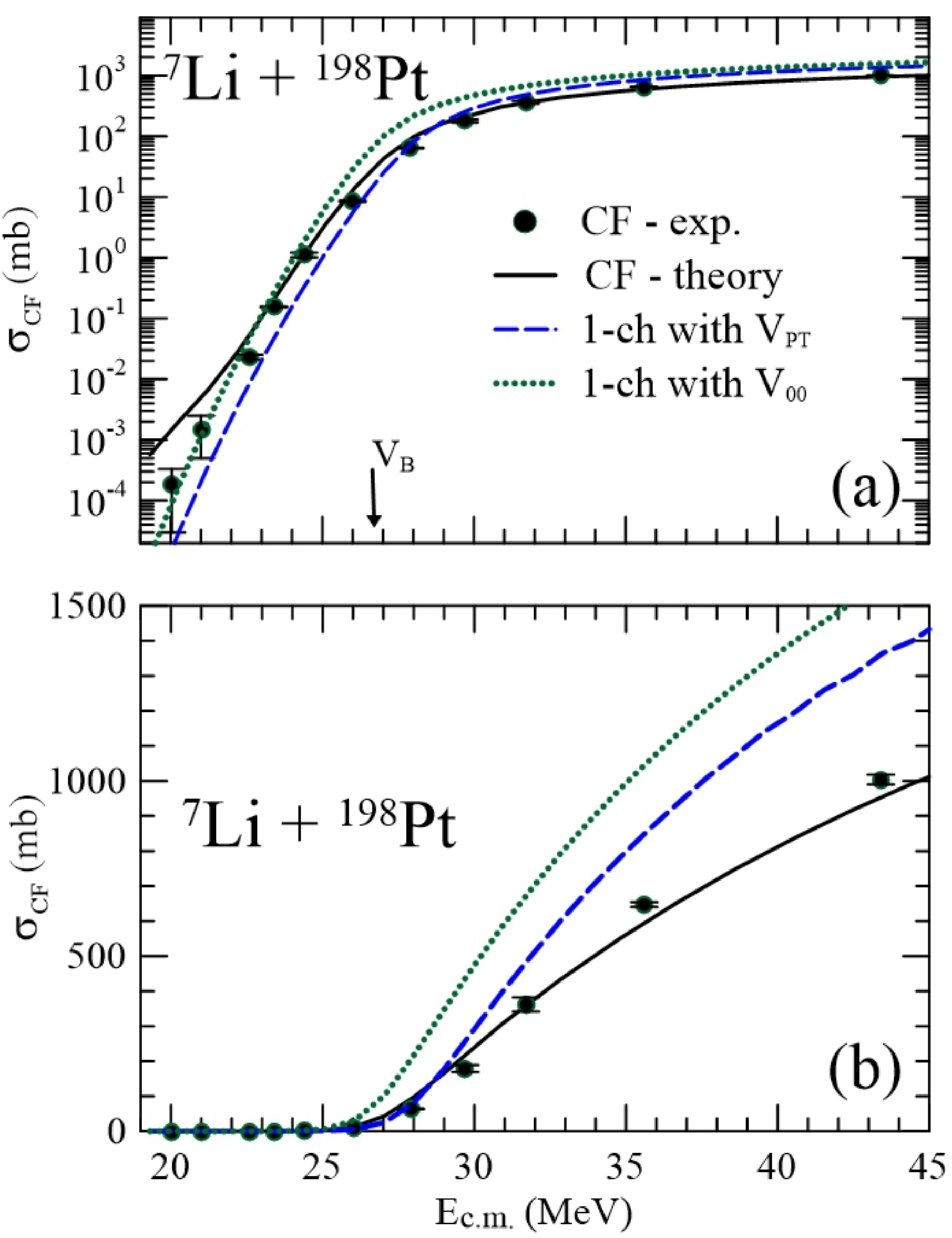}
\end{center}
\caption{(Color on line) Same as Fig.~\ref{F8}, but  for the $^7$Li + $^{198}$Pt system. Here the data are from Ref.~\cite{SND13}.}
\label{F11}
\end{figure}

We used our theory to calculate CF cross sections for collisions of $^7$Li projectiles with $^{209}$Bi, $^{197}$Au, $^{124}$Sn, and $^{198}$Pt targets. 
These targets have the advantage of not having excited states strongly coupled to the elastic channel. The results (solid black lines) are shown in 
Figs.~\ref{F8} to \ref{F11}. In each case, they are compared with the available  experimental data. All calculations were performed with the spectroscopic
amplitude $\mathcal{S} = 0.8$, which gave best results for the $^7$Li + $^{209}$Bi system. Note that the present results for this system are very close
to the ones presented in our previous work~\cite{RCL20}, but they are not exactly the same. This is due to the inclusion of the spectroscopic amplitude 
and to the use of a slightly improved mesh in the continuum discretization. \\

Figs.~\ref{F8} to \ref{F11} also show cross sections of two one-channel calculations. In the first (green dotted lines), we used the nuclear potential 
$V_{\scr 00}(R)$, which is obtained by folding the fragments-target interactions with the ground state density of the projectile (see Eq.~(\ref{V00})). 
In the second (blue dashed lines), we used the S\~ao Paulo potential between the projectile and the target, which ignores the cluster structure of $^7$Li
completely. Thus, the former 
takes into account the static effect of the low breakup threshold, whereas the latter does not. Both one-channel calculations were performed with typical 
short-range imaginary potentials, $W_{\scr PT}(R)$, given by WS functions with radii $R_0 = 1.0\,\left(A_{\scr P}^{\scr 1/3}\,+\,A_{\scr T}^{\scr 1/3}\right)$ fm, 
depth $W_0 = 100$ MeV and diffusivity $a = 0.2$ fm.\\

The overall agreement between the CF cross sections calculated by our method and the experimental data is quite good. The theoretical cross sections for 
the $^7{\rm Li} + ^{209}$Bi (Fig.~\ref{F8}),  $^7{\rm Li} + ^{197}$Au (Fig.~\ref{F9}), and $^7{\rm Li} + ^{124}$Sn (Fig.~\ref{F10}) systems are very close to
the data at all collision energies, above and below the Coulomb barrier. In the case of the $^{198}$Pt target (Fig.~\ref{F11}), the situation is not as good. 
The theoretical CF 
cross section is in excellent agreement with the data around and above the Coulomb barrier, but it overestimates the experimental results at energies 
well below $V_{\scr B}$.  In fact, this problem is not related to the target. It is a consequence of the extended energy range of the experiment~\cite{SND13}. 
It reaches energies $\sim 6$ MeV below the Coulomb barrier, where the cross sections are as low as $\sim 10^{-4}$ mb. The data for the other systems 
studied here are restricted to energies $E_{\rm c.m.} \gtrsim V_{\scr B} - 4\,{\rm MeV}$, where the cross sections are three orders of magnitude larger. \\

\begin{figure}
\begin{center}
\includegraphics*[width=7 cm]{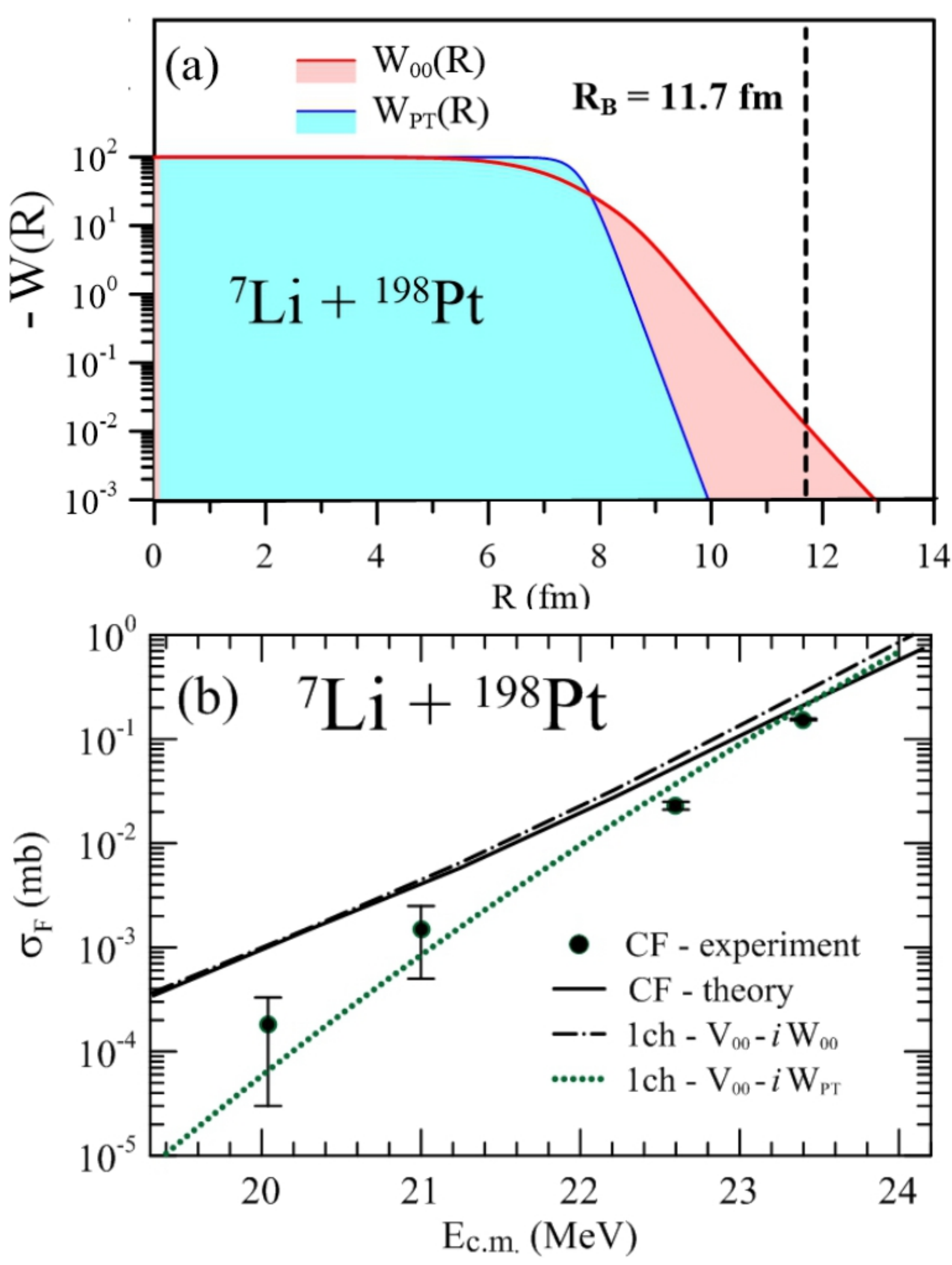}
\end{center}
\caption{(Color on line) (a) The imaginary potentials $W_{\scr 00}(R)$ and $W_{\scr PT}(R)$, shown in a logarithmic scale; (b) Fusion cross sections at very low energies. The 
CF cross section of our method (solid green line), and the fusion cross sections of one-channel calculations with the potentials $V_{\scr 00}(R) -i W_{\scr PT}(R)$ (red dotted line) 
and $V_{\scr 00}(R) -i W_{\scr 00}(R)$ (black dot-dashed line) are compared with the CF data of Refs.~\cite{SND13}. See the text for details.}
\label{F12}
\end{figure}
The inaccuracy of the theoretical CF cross section at energies well below $V_{\scr B}$ can be traced back to the imaginary potential, $W_{\scr 00}(R)$, used
in the CDCC calculations. This potencial, evaluated internally within the FRESCO code, is given by the expression
\begin{equation}
W_{\scr 00}(R) = \int d^3{\bf r}\, \left| \phi_0({\bf r})  \right|^2\ 
\left[  
\mathbb{W}^{\sc (1)}(r_1) \,+\, \mathbb{W}^{\sc (2)}(r_2) \right] ,
\label{W00}
\end{equation}
where $r_1$  and $r_2$ are the distances between the centres of the two fragments and the target. Although the ranges of the imaginary potentials $\mathbb{W}^{\sc (i)}$ are 
very short, the long tail of $\left| \phi_0({\bf r})  \right|^2$ extends $W_{\scr 00}(R)$ to large distances, beyond $R_{\scr B}$. This is illustrated in panel (a) of 
Fig.~\ref{F12}, which compares the imaginary potentials $W_{\scr 00}(R)$ and $W_{\scr PT}(R)$. Clearly, the tail of $W_{\scr 00}(R)$ has a considerably longer range.
This difference is not relevant at collision energies above $V_{\scr B}$, where the incident wave reaches the inner region of the barrier, where the two
imaginary potentials are very strong. In this case, the wave is strongly absorbed by both imaginary potentials. In this way, the fusion cross sections 
calculated with $W_{\scr 00}(R)$ and $W_{\scr PT}(R)$ are very close. The situation is different at very low collision energies, where the transmission
coefficient through the barrier is extremely small. Then, the cross section has a strong dependence on the tail of the imaginary potential, which, as shown in
the figure, is much longer for $W_{\scr 00}(R)$. However, this long-range absorption cannot be associated with fusion. Since the relevant direct channel, 
namely breakup, is explicitly included in the CDCC equations, this kind of absorption is spurious. It has no physical meaning. \\

A more quantitative picture of the problem is presented in panel (b) of Fig.~\ref{F12}, which shows the data of Refs.~\cite{DHH02,DGH04} at energies well 
below the Coulomb barrier, in comparison with different theoretical cross sections. The black solid line and the green dotted line are the same curves of  
Fig.~\ref{F11}. They represent, respectively, the CF cross section calculated by our method, and the one-channel cross section obtained with the complex 
potential $U = V_{\scr 00}-i \, W_{\scr PT}$. The third curve (black dot-dashed line) represents the results of a one-channel calculation  with the potential
$U = V_{\scr 00}-i \, W_{\scr 00}$. It corresponds to the limit of our CDCC calculation when all channel-couplings are switched off. The difference between the two 
one-channel calculations is the range of the imaginary potential. First, one notices that the CF cross section converges to the black dot-dashed line
at very low energies. This is not surprising, since the coupling matrix-elements become negligibly small in the low energy limit. On the other hand, at the 
lowest energies, these cross sections become much larger than the one calculated with $W_{\scr PT}$, which is in very good
agreement with the data. Therefore, one concludes that the inaccuracy of our CF cross section at energies well below $V_{\scr B}$ arises from
the spurious tail of the imaginary potential in the CDCC calculations. In principle, this shortcoming could be easily fixed by correcting the asymptotic behavior
of $W_{\scr 00}(R)$. However, this is not an easy task, since it would require internal modifications of the FRESCO code.

\subsubsection{ The static effect of the low breakup threshold }


As mentioned before, the low breakup threshold of $^7$Li affects the CF cross section in two ways. The first is a static effect, arising from the low energy binding 
the triton to the $\alpha$-particle, which leads to a long tail in the nuclear density. This makes the Coulomb barrier lower, enhancing fusion. On the other hand, 
the reaction dynamic is strongly affected by couplings with the breakup channel. This has a major influence on fusion, as will be demonstrated in the 
next sub-section.\\

\begin{table}
\centering
\caption{ Coulomb barriers of $V_{\scr PT}$ and $V_{\scr 00}$, for the systems studied in this work. The fourth column gives the barrier lowering in each 
case, and the fifth column is the ratio of the one-channel fusion cross sections calculated with the potentials $V_{\scr 00}$ and $V_{\scr PT}$, at 10 MeV 
above $V^{\scr 00}_{\scr B}$. See the text for details. }
\vspace{0.2cm}
\begin{tabular}{cccccccc}
\hline 
\ \ \ \ \ \ \ System\qquad\qquad\qquad  & \ \ $Z_{\scr T}$\ \ \ &\; $V^{\scr PT}_{\scr B}$\; &\; $V^{\scr 00}_{\scr B}$ \; & \;$\Delta V_{\scr B}$\ \ \ \  & $\mathcal{R}^{\scr 00}_{\scr PT}$\\ 
\hline                               
$^7$Li\,+\,$^{209}$Bi & 83 &29.36 & 28.29    & 1.07 & 1.21  \\
\,$^7$Li\,+\,$^{197}$Au & 79 & 28.25 & 27.21  & 1.04 & 1.20  \\
$^7$Li\,+\,$^{198}$Pt & 78 & 27.83 & 26.81   & 1.02 &  1.21  \\
$^7$Li\,+\,$^{124}$Sn & 50 &19.29 & 18.50  & 0.79 & 1.18  \\
\hline
\end{tabular}
\label{barriers}
\end{table}

\smallskip

Table \ref{barriers} shows Coulomb barriers associated with $V_{\scr PT}$ and  $V_{\scr 00}$, denoted respectively by $V^{\scr PT}_{\scr B}$ and
$V^{\scr 00}_{\scr B}$. As expected, the latter is systematically lower than the former. The reduction of the
barrier height increases with the charge of the target (or with the barrier height). For the systems studied in this work, it ranges from $\sim 0.8$ to 
$\sim 1.1$ MeV. The barrier lowering enhances the fusion cross section for the potential 
$V_{\scr 00}$, with respect to that for $V_{\scr PT}$. At $\sim 10$ MeV above the barrier, the ratio of the two cross sections for the four systems is
of the order of 1.2 or, more precisely, between 1.18 and 1.21.
%

\subsubsection{ CF suppressions at above-barrier energies } 


Now we compare the suppressions of CF for the different systems studied here. Since the cross sections depend on trivial factors, like the charges and 
sizes of the collision partners, direct comparisons of $\sigma_{\scr CF}$ do not give reliable information on reaction mechanisms. For a proper comparison,
one should first eliminate the influence of such undesirable factors. This is done through transformations on the cross sections and collision energies, known 
as {\it reduction procedures}. Several proposals can be found in the literature~\cite{CMG15}, but the most effective procedure for fusion data is the 
so called {\it fusion function} method~\cite{CGL09a,CGL09b}. It consists in the following transformations:
\begin{equation}\
E\ \longrightarrow \ x =\frac{E-V_{\scr B}}{\hbar\omega}, \qquad
\sigma_{\scr F} \ \longrightarrow\ F(x)=\frac{2E}{R_{\scr B}^2 \hbar\omega} \times \sigma_{\scr F}.
\label{FFtrans} 
\end{equation}
This method is based on the Wong's approximation~\cite{Won73} for the fusion cross section,
\begin{equation}\
\sigma _{\scr F}^{\scr W}=R_{\scr B}^2\  \frac{\hbar\omega}{2E}\, \ln \left[1 + \exp\left(2\pi\ \frac{E-V_{\scr B}}{\hbar\omega}  \right) \right].
\label{Wong}
\end{equation}
It can be immediately checked that if the fusion cross section is well approximated by Wong's formula, the fusion function takes the universal
form
\begin{equation}\
F_0(x)= \Big[1 + \exp\left(2\pi\,x \right) \Big].
\label{Wong-1}
\end{equation}
This expression was called the Universal Fusion Function (UFF) in Refs.~\cite{CGL09a,CGL09b}. Deviations from this behaviour are then 
associated with particular nuclear structure properties of the collision partners.\\

To carry out a comparative study of CF suppression at above-barrier energies, we apply the above prescription to collisions of $^7$Li with the 
$^{209}$Bi, $^{197}$Au, $^{124}$Sn, and $^{198}$Pt targets. We consider both the theoretical and experimental CF cross sections, discussed in
the previous sub-sections. The results are denoted by $F_{\rm th}(x)$ and $F_{\rm exp}(x)$, respectively. Further, there are two possibilities.
The transformations of Eq.~(\ref{FFtrans})  can be based on the barrier parameters of the potential $V_{\scr 00}$ ($V_{\scr B}^{\scr 00}, R_{\scr B}^{\scr 00}$ and 
$\hbar\omega^{\scr 00}$), or on the parameters of $V_{\scr PT}$ ($V_{\scr B}^{\scr PT}, R_{\scr B}^{\scr PT}$ and $\hbar\omega^{\scr PT}$). In this way,
one can evaluate two theoretical fusion functions, $F_{\rm th}^{\scr 00}(x)$ and $F_{\rm th}^{\scr PT}(x)$, and two experimental fusion functions,
$F_{\rm exp}^{\scr 00}(x)$ and $F_{\rm exp}^{\scr PT}(x)$. Note that the fusion functions $F^{\scr 00}(x)$ and $F^{\scr PT}(x)$ have very
different meanings, as discussed below.\\

In the present work, the investigated nuclear structure property is the low breakup threshold of $^7$Li. Since we chose targets that do 
not have excited states strongly coupled to the elastic channel, the CF fusion functions may be directly compared with the UFF. As the potential 
$V_{\scr PT}$ completely ignores the cluster structure of the projectile and its binding energy, comparisons of $F_{\rm th}^{\scr PT}(x)$ and of 
$F_{\rm exp}^{\scr PT}(x)$ with the UFF give the global influence of the low binding on the theoretical and on the experimental CF cross sections, 
respectively. That is, they measure the net result of the competition between the barrier lowering enhancement and breakup coupling suppression 
on CF. On the other hand, comparisons of $F_{\rm th}^{\scr 00}$ and 
$F_{\rm exp}^{\scr 00}$ with the UFF give a different piece of information. Since $V_{\scr 00}$ takes into account the long tail of the $^7$Li density, the static 
effects associated with the barrier lowering are cancelled in these fusion functions. Therefore, their comparisons with the UFF measure exclusively the influence
of couplings with the breakup channel. \\

\begin{figure}
\begin{center}
\includegraphics*[width=7 cm]{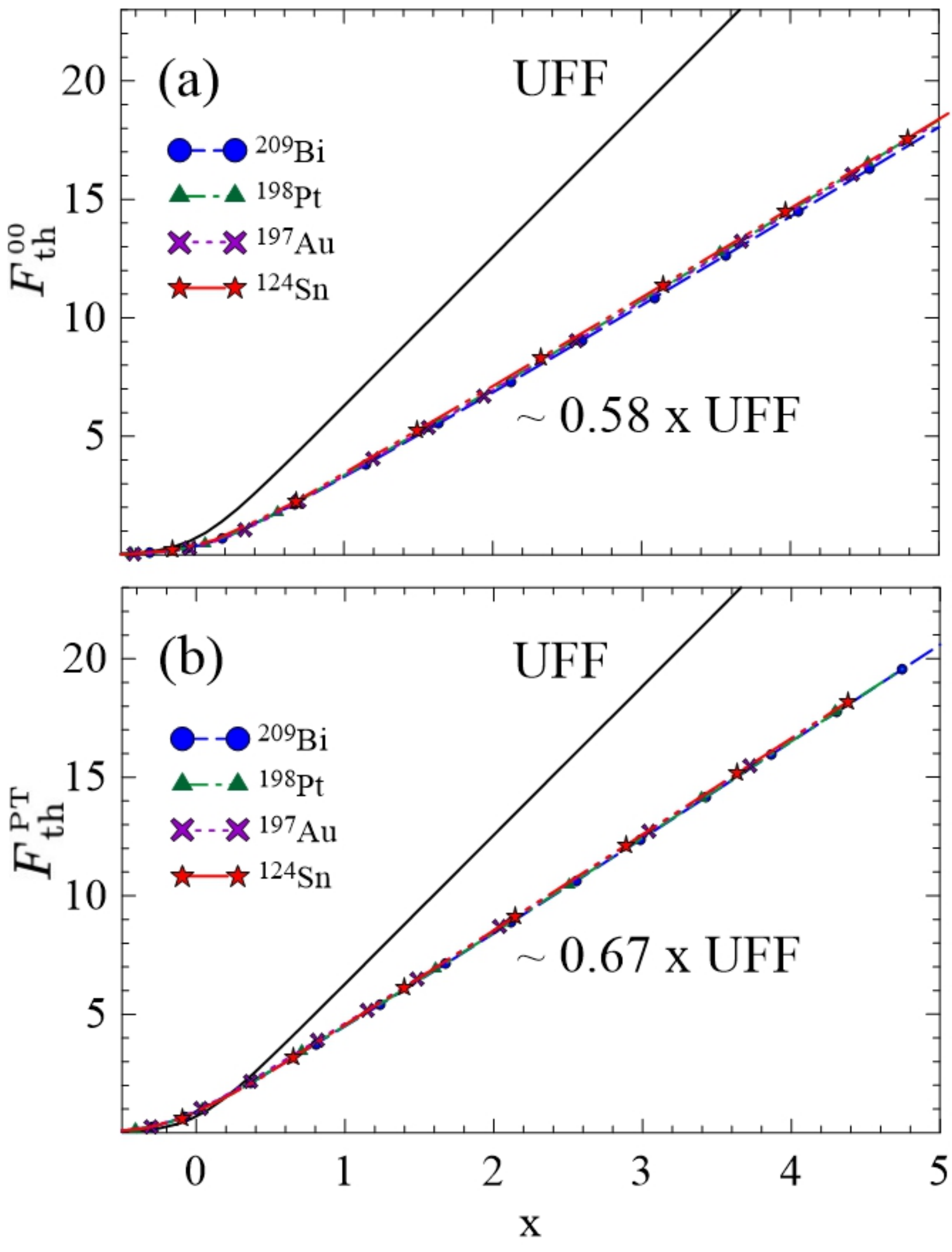}
\end{center}
\caption{(Color on line) Theoretical fusion functions $F_{\rm th}^{\scr 00}$ (panel (a)) and $F_{\rm th}^{\scr PT}$ (panel (b)) in collisions of $^7$Li with the
$^{209}$Bi, $^{197}$Au, $^{124}$Sn and $^{198}$Pt targets. See the text for details.}
\label{F13}
\end{figure}

Fig.~\ref{F13} shows the theoretical fusion functions for the systems studied here. Since we are interested in the suppression at above-barrier energies, the 
plots are shown only in a linear scale. First, one notices that both the $F_{\rm th}^{\scr 00}$ and $F_{\rm th}^{\scr PT}$ fusion functions are nearly system 
independent. The lines for the different targets can hardly be distinguished from each other. To very good approximations, one can write: 
\begin{equation}
F_{\rm th}^{\scr 00}(x) \simeq 0.67 \times F_0(x);\ \ \ F_{\rm th}^{\scr PT}(x) \simeq 0.58 \times F_0(x),
\label{factors}
\end{equation} 
where $F_0(x)$ is the universal fusion function of Eq.~(\ref{Wong-1}).
The above equation indicates that $F_{\rm th}^{\scr 00}$ and $F_{\rm th}^{\scr PT}$ are suppressed with respect to the UFF by 33 and 42\%, respectively.\\

\begin{figure}
\begin{center}
\includegraphics*[width=7 cm]{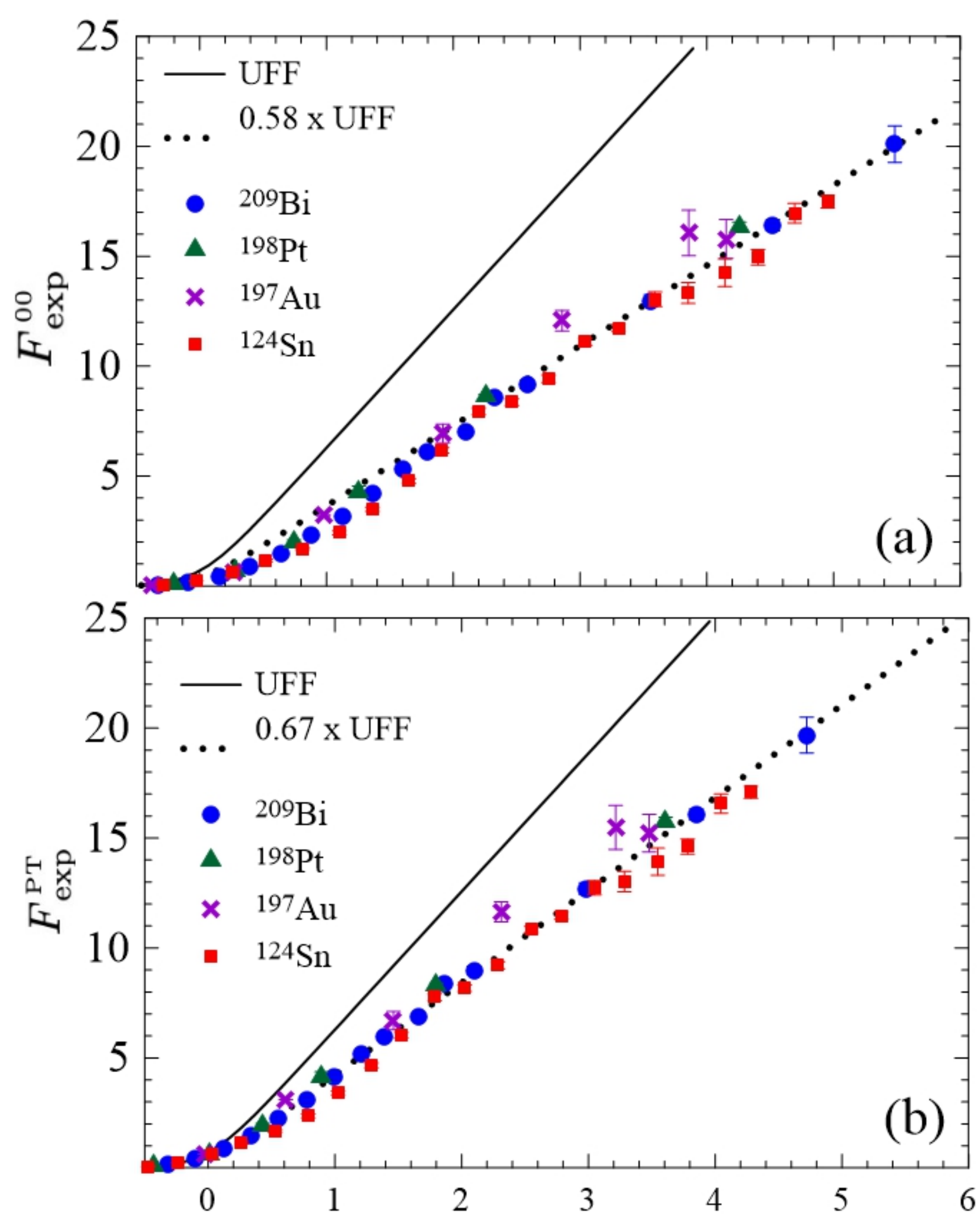}
\end{center}
\caption{(Color on line) Experimental fusion functions $F_{\rm exp}^{\scr 00}$ (panel (a)) and   $F_{\rm exp}^{\scr PT}$ (panel (b)) in collisions of $^7$Li with the
$^{209}$Bi, $^{197}$Au, $^{124}$Sn and $^{198}$Pt targets. See the text for details.}
\label{F14}
\end{figure}

Fig.~\ref{F14} shows the experimental fusion functions corresponding to the theoretical curves of the previous figure. The dotted lines represent
the predictions of our theory for the two fusion functions, within the $0.67 \times F_0(x)$ and $0.58\times F_0(x)$ approximations. Clearly the data
follow very closely the behaviour predicted by the theory, except por a few data points that present small fluctuations around the dotted lines. 
Usually, CF suppression factors are obtained comparing the data with predictions of barrier penetration models (or results of one-channel calculations), 
based on projectile-target potentials that ignore the low breakup threshold. Thus, they should be compared with suppression factors extracted from 
$F_{\rm exp}^{\scr PT}$. Dasgupta {\it et al.}~\cite{DHH02,DGH04} studied the $^7$Li + $^{209}$Bi system, and found a ratio of 0.74 between the CF
data and predictions of barrier penetration models. This is a bit larger than the 0.67 factor, appearing in Fig.~\ref{F14}.  The difference can be traced back 
to the different potential used by these authors in their barrier penetration model calculation. They adopted the Aky\"uz-Winther (AW) potential, instead 
of the SPP used in the present work. The barrier for the AW potential is 0.4 MeV higher than that for the SPP~\cite{CGL14} and, consequently, the cross sections obtained 
with the former is $\sim 10\%$ lower than that of the SPP. Taking this difference into account, our suppression factor becomes very close to theirs.

 
 \subsection{Incomplete fusion cross sections}\label{sect ICF}
 
 
\subsubsection{$^7$Li + $^{209}$Bi}

\begin{figure}
\begin{center}
\includegraphics*[width=7 cm]{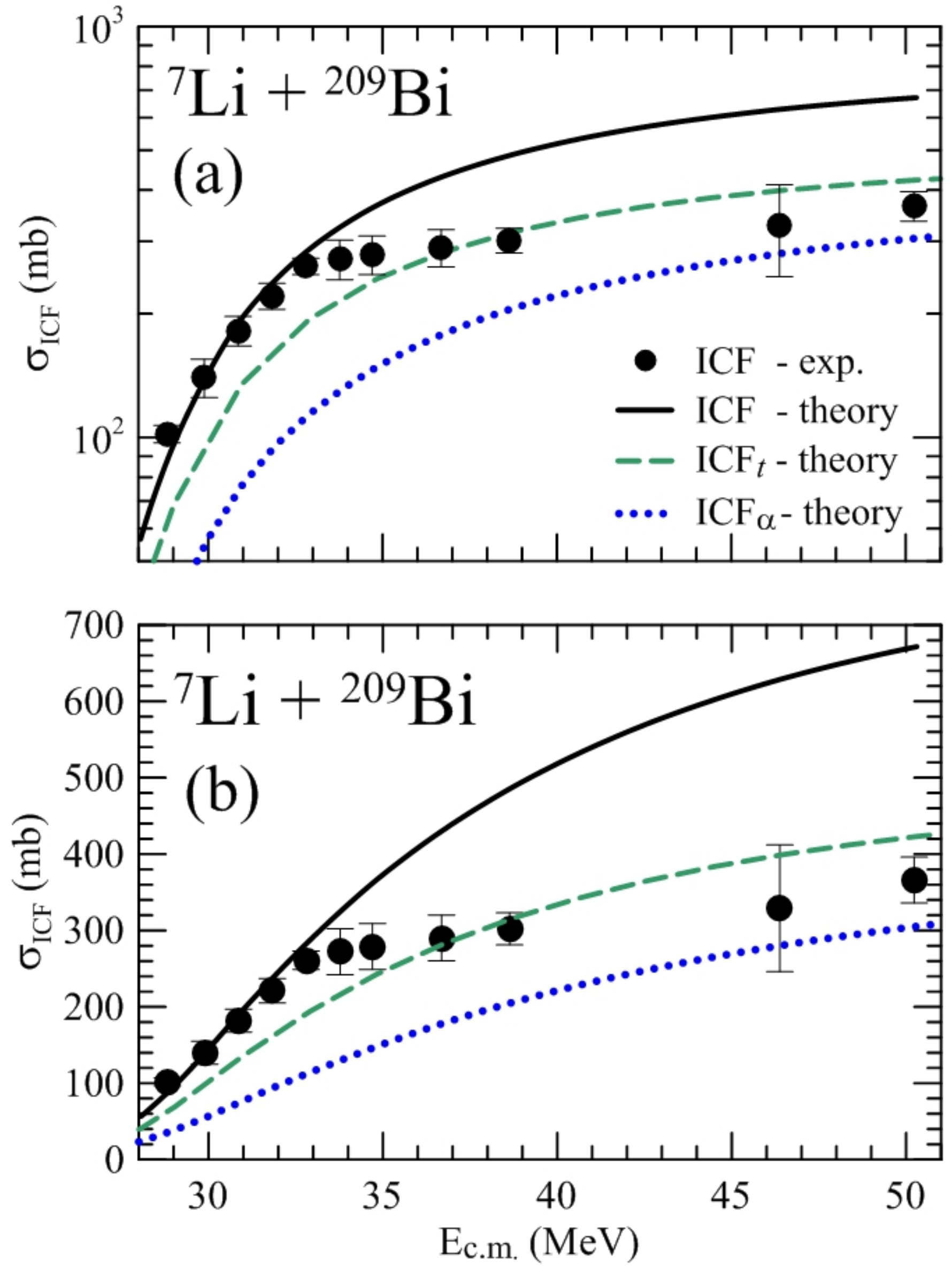}
\end{center}
\caption{(Color on line) Incomplete fusion cross sections for the $^7$Li + $^{209}$Bi system calculated by our method, in comparison with the ICF data of 
Refs.~\cite{DHH02,DGH04}.}
\label{F15}
\end{figure}

Fig.~\ref{F15} shows ICF cross sections for the $^7{\rm Li}\,+\,^{209}{\rm Bi}$ system calculated by the method of the present work. The cross section 
for the triton (${\rm ICF}_t$) and $\alpha$-particle (${\rm ICF}_\alpha$) captures are represented, respectively, by a green dashed line and a blue dotted 
line. The solid black line corresponds to the full ICF cross section, namely $\sigma_{\scr ICF} =  \sigma_{{\scr ICF}_{\scs t}} + \sigma_{\scr ICF_\alpha}$. Our 
results are compared to the ICF data of Dasgupta {\it et al.}~\cite{DHH02,DGH04}, obtained detecting characteristic $\alpha$-particles. Note that this 
experiment could not distinguish the ICF$_t$ and the ICF$_\alpha$ components of $\sigma_{\scr ICF}$. To clarify the situation,
we give some details of this work. The ICF$_t$ process leads to the formation of $^{212}$Po, and the lighter $^{211,210,209}$Po isotopes, 
through successive neutron emissions. On the other hand, ICF$_\alpha$ produces $^{213}$At and other lighter isotopes by neutron evaporation. In both cases,
the Po and the At isotopes de-excite by $\alpha$-decay. The emitted $\alpha$-particles are detected, and the parent nuclei are identified by their energies and  
half-lives. In principle, this procedure could lead to the individual ICF$_t$ and ICF$_\alpha$ cross section. However, $^{210}$At decays almost completely by 
$\beta^{\scr +} + {\rm EC}$ to $^{210}$Po. In this way, the At and the Po decay chains are mixed. Thus, an $\alpha$-particle emitted by $^{210}$Po is 
a signature of ICF, but one cannot tell whether it is ICF$_t$ or ICF$_\alpha$. For this reason, this experiment determines only their sum, $\sigma_{\scr ICF}$. 
By inspecting Fig.~\ref{F15}, we find that the predictions of our theory at low energies are very accurate. The four data points at the lowest  energies fall on top of 
the theoretical curve. However, the calculated cross section above $\sim 35$ MeV overestimates the data. It grows continuously with the energy, whereas 
the data are roughly constant. Nevertheless, the discrepancy between theory and experiment might, at least in part, arise from missing contributions from 
$^{209}$Po, in the decay chains of both ICF processes. Owing to its long half-life ($\sim 100$ y), its $\alpha$-decay could not be measured. 
Dasgupta {\it et al.}~\cite{DHH02,DGH04} estimated the contribution from this channel using the PACE evaporation code~\cite{Gav80}. They found that it 
should be negligible at the lowest energies of the experiment, but it becomes important above $\sim 36$ MeV. For this reason, they suggested that the data
above this limit should be considered as a lower bound to the actual cross section. Thus, our results may be consistent with the data in this energy range.

Finally, comparing the theoretical ICF$_t$ and  ICF$_\alpha$ cross sections, we conclude that the ICF$_t$ component of $\sigma_{\scr ICF}$ is dominant, 
but the ${\rm ICF_\alpha}$ component is appreciable. At above barrier energies, $\sigma_{\scr ICF_\alpha}$ is about 50\% of $\sigma_{{\scr ICF}_{\scs t}}$.

 
\subsubsection{$^7$Li + $^{197}$Au}


%
\begin{figure}
\begin{center}
\includegraphics*[width=7 cm]{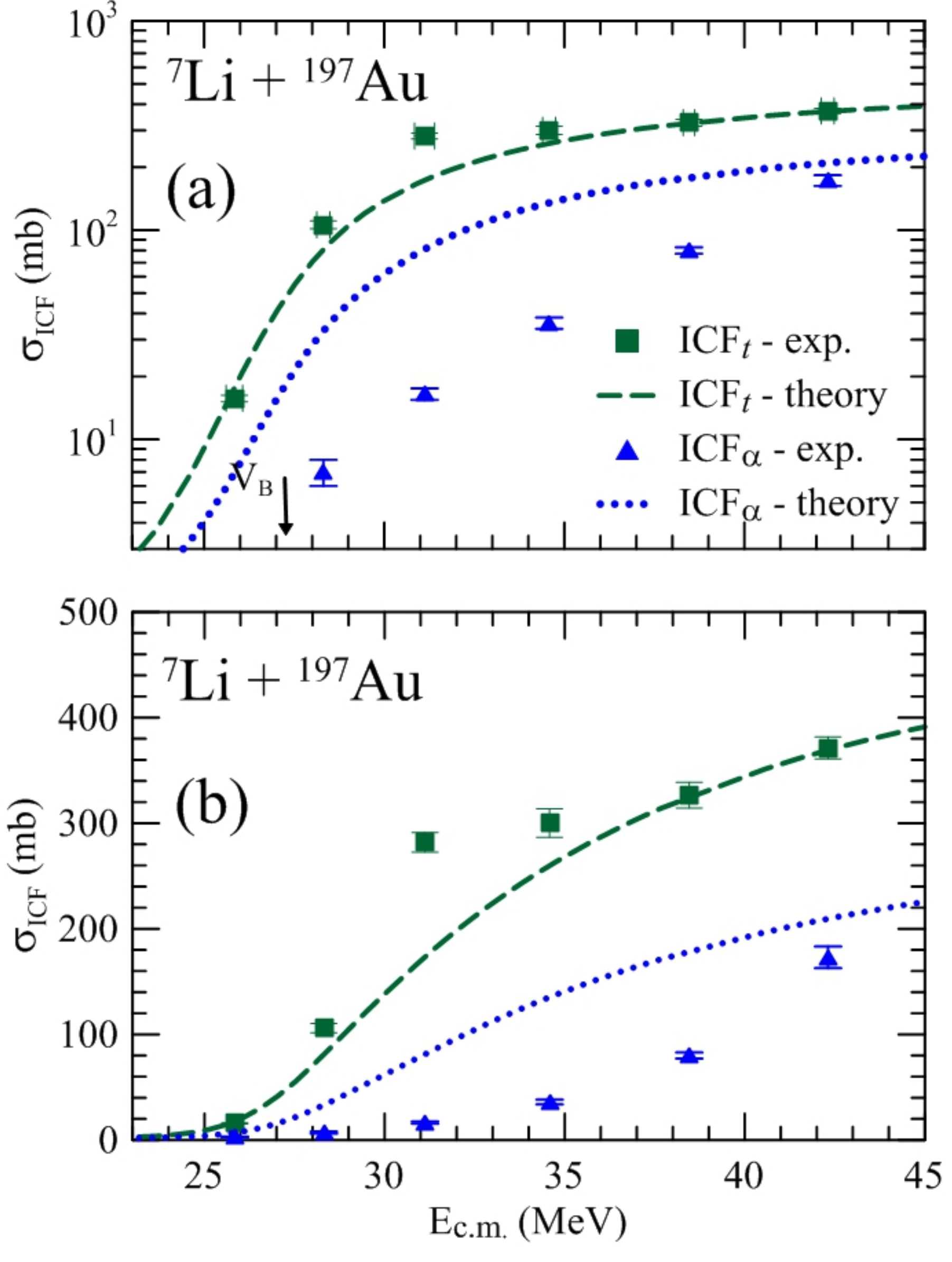}
\end{center}
\caption{(Color on line) ICF$_{\rm t}$ and ICF$_{ \alpha}$ cross sections for the $^7$Li + $^{197}$Au system calculated by our method, in
comparison to the data of Palshetkar {\it et al.}~\cite{PTN14,nndc}.}
\label{F16}
\end{figure}
Fig. \ref{F16} shows the $\sigma_{{\scr ICF}_{\scs t}}$ (green dashed line) and $\sigma_{\scr ICF_\alpha}$ (blue dotted line) cross sections for the 
$^7{\rm Li}\,+\, ^{197}{\rm Au}$ system, calculated by our method. The results are compared to the experimental cross sections of 
Palshetkar {\it et al.}~\cite{PTN14,nndc}, measured by the gamma-ray spectroscopy method (in- and off-beam). Note that in this experiment, it was possible
to determine individual cross sections for each ICF process. Inspecting the figure, we conclude that the $\sigma_{{\scr ICF}_{\scs t}}$ cross section predicted by 
our method reproduces very well the data, except for the data point at $E_{\rm c.m.}\simeq 31$ MeV, which is $\sim 30\%$ larger than the theoretical prediction. \\

On the other hand, the theoretical predictions for $\sigma_{\scr ICF_\alpha}$ are well above the data, except for the data point at the highest energy, where
the difference between the two cross sections is small. Note that the $\sigma_{\scr ICF_\alpha} / \sigma_{{\scr ICF}_{\scs t}}$ ratio at above-barrier energies 
predicted by our method is of the order of 50\%, similarly to the $^7$Li + $^{209}$Bi system.  The origin of the discrepancy between our predictions for 
$\sigma_{\scr ICF_\alpha}$ and the data is not clear to us. It calls for further investigations.

 
\subsubsection{$^7$Li + $^{124}$Sn}


\begin{figure}
\begin{center}
\includegraphics*[width=7 cm]{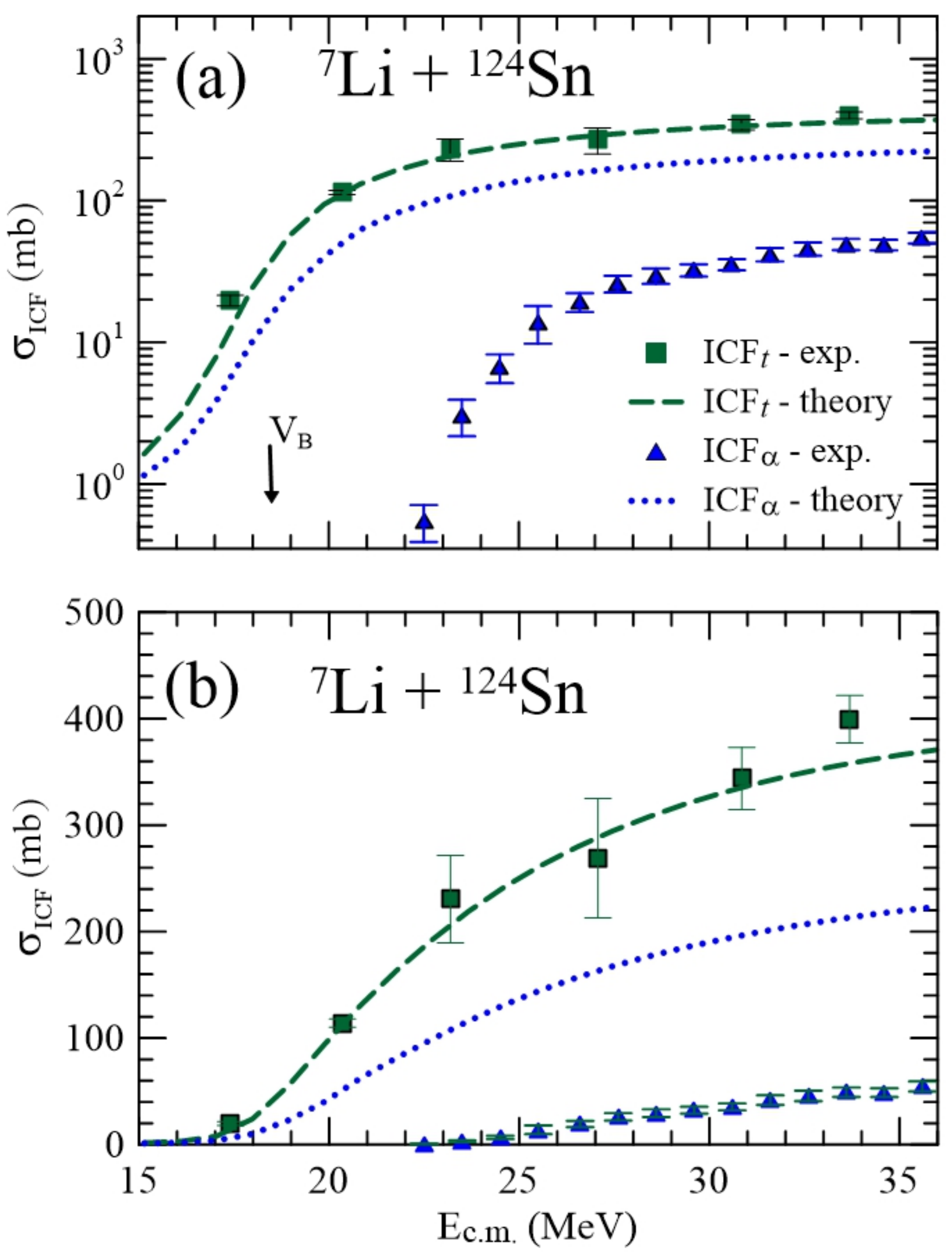}
\end{center}
\caption{(Color on line) Same as the previous figure, but now the system is $^7$Li + $^{124}$Sn, and the data are from Parkar {\it et al.}~\cite{PSP18}.}
\label{F17}
\end{figure}
Fig. \ref{F17} shows $\sigma_{{\scr ICF}_{\scs t}}$ and $\sigma_{\scr ICF_\alpha}$ cross sections calculated by our method for the $^7{\rm Li}\,+\, ^{124}{\rm Sn}$
system. The notation of the curves is the same as in the previous figure. Our results are compared to the experimental $\sigma_{{\scr ICF}_{\scs t}}$ and 
$\sigma_{\scr ICF_\alpha}$ cross sections of Parkar {\it et al.}~\cite{PSP18}, also measured by the gamma-ray spectroscopy method (in- and off-beam). The situation 
is very similar to that observed for the previous system. The $\sigma_{{\scr ICF}_{\scs t}}$ cross section predicted by our method is in excellent agreement with
the data, whereas our predictions for $\sigma_{\scr ICF_\alpha}$ are much larger than the data. At the highest energies of the experiment,
the theoretical $\sigma_{\scr ICF_\alpha} / \sigma_{{\scr ICF}_{\scs t}}$ ratio is slightly above 50\%, while the experimental ratio is of the order of 10\%.

 
\subsubsection{$^7$Li + $^{198}$Pt}


%
\begin{figure}
\begin{center}
\includegraphics*[width=7 cm]{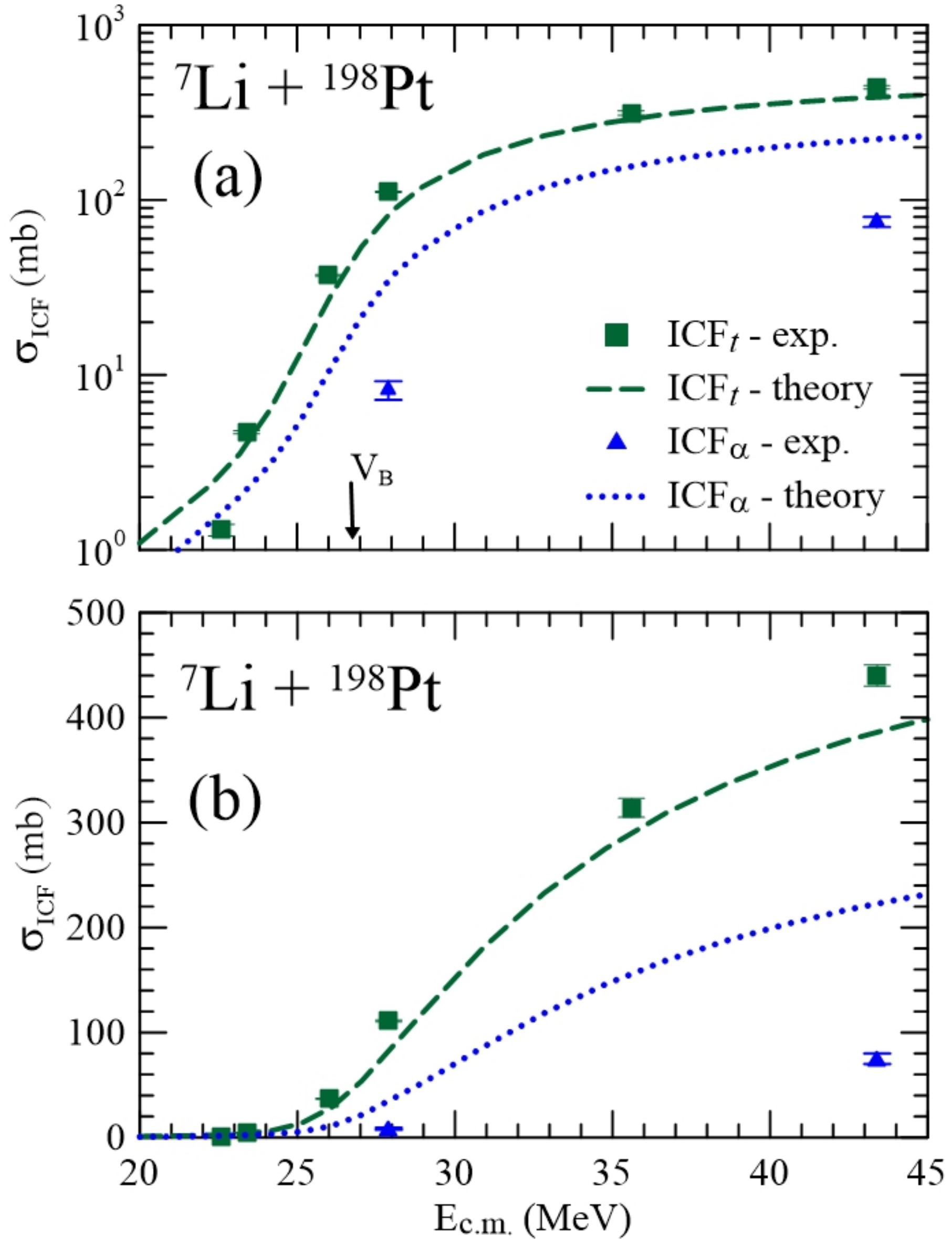}
\end{center}
\caption{(Color on line) Same as the previous figure, but now the system is $^7$Li + $^{194}$Pt, and the data are from Shrivastava {\it et al.}\cite{SND13}.}
\label{F18}
\end{figure}
Fig.~\ref{F18} shows $\sigma_{{\scr ICF}_{\scs t}}$ and $\sigma_{\scr ICF_\alpha}$ cross sections calculated by our method for the 
$^7{\rm Li}\,+\, ^{198}{\rm Pt}$ system, in comparison with the data of Shrivastava {\it et al.}~\cite{SND13}. Again, the experiment used the gamma-ray spectroscopy
method and was able to measure individual cross sections for the two ICF processes. The situation is similar to those observed for the $^{197}$Au and $^{124}$Sn
targets. The theoretical predictions for $\sigma_{{\scr ICF}_{\scs t}}$ are very close to the data, whereas those for $\sigma_{\scr ICF_\alpha}$ overpredict
them. However, here there is a difference. As in the case of CF, the theoretical cross section at the lowest data point is much larger than the data. This
problem is related to the overprediction of CF at very low energies for the system. We believe that it arises from the long tail of the imaginary potential in the 
CDCC calculations but this requires further investigation.


\section{Conclusions} 


We  gave a detailed presentation of the new method introduced in a previous paper~\cite{RCL20}, to evaluate CF and ICF cross sections in collisions of weakly 
bound projectiles. Our method has the advantages of fully accounting for the influence of continuum wave functions on the fusion processes,  and of being 
applicable to any weakly bound projectile that breaks up into two fragments. The method was used to evaluate CF and ICF cross sections in collisions of $^7$Li 
with several targets, and the results were compared with the available data. \\

At near-barrier and above-barrier energies, the agreement between our theoretical CF cross section and the data is excellent. However, at energies well
below the Coulomb barrier, our cross section overestimates the data. We have shown that this is a consequence of the long tail of the imaginary potential
evaluated within the FRESCO code. In this energy region, this tail leads to absorption beyond the radius of the Coulomb barrier, which does not represent fusion.
This problem is more serious in collisions of projectiles with lower binding energies, like $^6$Li, and this situation is still much worse for projectiles far from stability, 
like $^8$B or $^{11}$Li. Presently, a correction of this problem is under work.\\

The situation for ICF cross sections is more complex. In the case of the $^7{\rm Li}\,+\,^{209}{\rm Bi}$ system, our ICF cross section was compared with the 
experimental results of Dasgupta {\it et al.}, obtained through alpha-particle measurements. At low energies, the agreement between theory and experiment
is excellent. At $E_{\rm c.m.} \gtrsim 35$ MeV, the theoretical cross section overpredicts the data, but this may be due, at least in part, to missing contributions 
from the long-lived $^{209}$Po isotope, which becomes important in this energy region. The theoretical ICF cross sections for the $^{197}$Au, $^{124}$Sn,
and $^{198}$Pt targets were compared with experimental cross sections measured by the gamma-ray spectroscopy method (in- and off-beam). In this
case, there are individual data for the ${\rm ICF}_{\scs t}$ and ${\rm ICF_\alpha}$ processes. We found that our theory reproduces the ${\rm ICF}_{\scs t}$ data 
with high accuracy, but it systematically overpredicts $\sigma{\scr ICF_\alpha}$. This discrepancy deserves further investigations. \\

The method of the present work can be extended in several directions. One could, for example, include target excitations or even study collisions of projectiles
like $^9$Be or $^{11}$Li, which break up into three fragments.  Modifying our code to handle these problems would be straightforward. However, it uses
radial wave functions extracted from FRESCO. Then, it would be necessary to modify the form factors in the CDCC equations, so as to include the influence of the
new degrees of freedom. This is a hard task because the form factors are evaluated within the FRESCO. The implementations of these extensions are in progress.


\appendix

\section{Calculation of the absorption probability}


In this appendix we evaluate the probabilities $P_{\scr B}^{\scr (i)}(J)$ and $P_{\scr C}^{\scr (i)}(J)$ of Sect.~\ref{theory}. We consider the collision of a projectile formed by two fragments,
one with spin zero and the other with $s$, on a spinless target. In this case, the contribution from the absorption of fragment $c_i$ to the TF cross section is given by the expression
\begin{equation}
\sigma^{\scr (i)}_{\scr TF} = \frac{K}{E}\ \frac{(2\pi)^3}{\left( 2j_0+1 \right)}\ \sum_{\nu_0}\big< \Psi^{\scr (+)}_{\scr \bf{k}\, j_o \nu_o} \big|\, \mathbb{W}^{\scr (i)}
\, \big| 
\Psi^{\scr (+)}_{\scr \bf{k}\, j_o \nu_o} \big> ,
\label{sigf0}
\end{equation}
where $\Psi^{\scr (+)}_{\scr \bf{k}\, j_o \nu_o}$ is the scattering wave function for a collision with wave vector ${\bf k}$, initiated with intrinsic angular momentum $j_0$ and
z-component $\nu_0$. In this equation, the normalization constant of Eq.~(\ref{sigTF-1}) was set as $A=(2\pi)^{\scr -3/2}$.\\

The angular momentum projected scattering wave function is obtained coupling the intrinsic angular momentum  (${\bf j}_\alpha$) with the orbital angular momentum of the 
projectile-target motion (${\bf L}$).  It is given by~\cite{CaH13} ,
\begin{multline}
\Psi^{\scr (+)}_{\scr \mathbf{k}\, j_o \nu_o}({\bf R}, {\bf r})  = \frac{1}{(2\pi)^{\scr 3/2}} \
\sum_{\alpha J L L_0}  \,\ \frac
{{\mathcal U}_{\alpha  L, 0  L_0}^{J}(K_{\alpha},R)}{K R} \  e^{i\sigma_{\scr L_0}}\\
\times \sqrt{4\pi\ (2L_0+1)}\ \big<  J \nu_0\, \big|  L_0\,0\, j_{0}\,\nu_0 \big>\  
\times 
\ \mathcal{Y}_{\sc \alpha L}^{\sc J \nu_0} \left( \hat{\bf R}, \zeta \right)  ,
\label{8Psi}
\end{multline}
where ${\mathcal U}_{\alpha  L, 0  L_0}^{J}(k_{\alpha},R)$ are the solutions of the radial equation and  $\mathcal{Y}_{\sc \alpha L}^{\sc JM} ( \hat{\bf R}, \zeta )$ are the spin-channel wave
functions (in the present case, the intrinsic coordinates, $\zeta$, are simply the components of the vector ${\bf r}$), 
\begin{multline}
\mathcal{Y}_{\sc \alpha L}^{\sc J\nu_0} ( \hat{\bf R}, \zeta) =   i^L\ 
\sum_{\nu M_{\sc L}}
\big< L\,M_{\sc L}\, j_\alpha\,  \nu \big| J\,\nu_0  \big> \\ 
\times Y_{L M_{\sc L}} (\hat{\bf R}) \
\phi_{\alpha j_\alpha   \nu}(\zeta) .
\label{spin-chan}
\end{multline}
Above, $\phi_{\alpha j_\alpha   \nu}(\zeta)$ is the eigenstate of the intrinsic Hamiltonian of the projectile with energy $\varepsilon_\alpha$, angular momentum $j_\alpha$ and projection
$\nu$ (the explicit form of these states will be discussed later).\\

Next, we carry out the multipole expansion of the imaginary potential,
\begin{equation}
\mathbb{W}^{\scr (i)}({\bf R},{\bf r}^\prime_i) =4\pi\   \sum_{\lambda,\mu}  \ (-)^\mu\ Y_{\lambda \mu}(\hat{\bf R})\, Q^{{\scr (i)}}_{\lambda \,-\mu}({\bf r}^\prime_i) .
\label{expW-1-a}
\end{equation}
where $Q^{{\scr (i)}}_{\lambda \,-\mu}({\bf r}^\prime_i) $ is the spherical tensor operator
\begin{equation}
Q^{{\scr (i)}}_{\lambda \,-\mu}({\bf r}^\prime_i) =  {\mathcal W}^{\sc (i)\lambda}(R,r^\prime_i)\  Y_{\lambda \, -\mu}(\hat{\bf r}^\prime_i) .
\label{expW-1-b}
\end{equation}

\bigskip

Using Eqs.~(\ref{8Psi}-\ref{expW-1-a}) in Eq.~(\ref{sigf0}), $\sigma^{\scs (i)}_{\scr TF}$ can be put in the form,
\begin{equation}
\sigma^{\scs (i)}_{\scr TF} = \frac{\pi}{K^2}\ \sum_J (2J+1)\  \, {\mathcal P}^{\scr (i)}(J)  ,
\label{TJ-1}
\end{equation}
where $P^{\scr (i)}(J)$ is the probability of absorption of fragment $c_i$ by the target in a collision with angular momentum  $J$, given by
\begin{multline}
P^{\scr (i)}(J) = \frac{4K}{E\,\hat{j}_0^2\,\hat{J}^2}\ \sum_\lambda \, \sum_{\alpha L L_0}\,  \sum_{\alpha^\prime L^\prime L^\prime_0}\,
(i)^{\scr L^\prime - L}\  \hat{L_0}\, \hat{L^\prime_0}\\
e^{ i  \left( \sigma_{\scr L_0^\prime} -  \sigma_{\scr L_0} \right)}
\times \sum_{\nu_0}\ \left\langle L_0 0 \j_0\nu_0     \big| J\nu_0\right\rangle\ \left\langle  J\nu_0  \big| L_0^\prime 0 \j_0\nu_0 \right\rangle\\
\times \int dR \ {\mathcal U}^{J*}_{\alpha  L, 0  L_0}(K_{\alpha} R) \ {\mathcal U}^{J}_{\alpha^\prime  L, 0  L_0^\prime}(K_{\alpha^\prime} R)\ X^{J{\scr (\lambda)}}_{\alpha L,\alpha^\prime L^\prime}(R).
\label{TFi-2}
\end{multline}
Above, we denote: $\hat{j}_0 =\sqrt{2j_0+1}$, and use an analogous notation for other angular momentum quantum numbers, and
\begin{equation}
X^{J{\scr (\lambda)}}_{\alpha L,\alpha^\prime L^\prime}(R) = 4\pi\ \left( \mathcal{Y}^{J\nu_0}_{\alpha L} \left| {\bf Y}_\lambda \cdot {\bf Q}_\lambda \right| 
\mathcal{Y}^{J\nu_0}_{\alpha^\prime L^\prime} \right).
\label{X-1}
\end{equation}
The above quantity seems to depend on $\nu_0$ but it actually does not. It cannot depend on orientation because $ {\bf Y}_\lambda \cdot {\bf Q}_\lambda$ is
a scalar. Thus, the $\nu_0$-dependence is restricted to the Clebsh-Gordan coefficients. Then, carrying out the sum over $\nu_0$, we get~\cite{BrS94}, 
\begin{multline}
\sum_{\nu_0}\ \left\langle L_0 0 \j_0\nu_0     \big| J\nu_0\right\rangle\ \left\langle  J\nu_0  \big| L_0^\prime 0 \j_0\nu_0 \right\rangle = \\
\hat{J}^2 \ 
\sum_{\nu_0}
\left(
\begin{array}
[c]{ccc}%
L_0\, &j_0 & J\\
0  & \nu_0 & -\nu_0
\end{array} 
\right)\, 
\left(
\begin{array}
[c]{ccc}%
L^\prime_0\, &j_0 & J\\
0  & \nu_0 & -\nu_0
\end{array} 
\right) =\\
\frac{\hat{J}^2}{\hat{L}_0^2}\ \delta_{L_0 L_0^\prime}.
\end{multline}
Using this result, Eq.~(\ref{TFi-2}) takes the form,
\begin{multline}
P^{\scr (i)}(J) = \frac{4K}{E\,\hat{j}_0^2}\ \sum_\lambda \, \sum_{\alpha L \alpha^\prime L^\prime\,  L_0 }\, (i)^{\scr L^\prime - L} 
 \int dR \ X^{{\scr (i)\lambda}}_{\alpha L,\alpha^\prime L^\prime}(R) \\
 \times {\mathcal U}^{J*}_{\alpha  L, 0  L_0}(K_{\alpha} R) \ {\mathcal U}^{J}_{\alpha^\prime  L, 0  L_0^\prime}(K_{\alpha^\prime} R)\ .
\label{TFi-3}
\end{multline}
%

\subsection{Evaluation of $X^{{\scr (i)\lambda}}_{\alpha L,\alpha^\prime L^\prime}(R)$}


Using the notation of Ref.~\cite{BrS94} for the wave functions: $\mathcal{Y}^{J\nu_0}_{\alpha L} \rightarrow \left| \alpha (L j_\alpha) J \right)$, Eq.~(\ref{X-1}) reads,
\begin{multline}
X^{{\scr (i)\lambda}}_{\alpha L,\alpha^\prime L^\prime}(R) =
 4\pi\  \big<\alpha (L j_\alpha) J \,\big|\!\big|\, 
{\bf Y}_\lambda(\hat{\bf R}) \\
\cdot {\bf Q}_\lambda(R,\zeta)\,\big|\!\big|\, \alpha^\prime (L^\prime j_{\alpha^\prime})  J\big> ,
\end{multline}
or (Eq.~(5.13) of Ref.~\cite{BrS94})
\begin{multline}
X^{{\scr (i)\lambda}}_{\alpha L,\alpha^\prime L^\prime}(R)  = 4\pi\ (-)^{\sc J-L- j_{\alpha^\prime}} \ \hat{L}\ \, \hat{j}_{\alpha} \ 
W(L L^\prime j_\alpha j_{\alpha^\prime} ;\lambda J)\\
 \times
\big< L \,\big|\!\big|\, {\bf Y}_\lambda(\hat{\bf R}) \,\big|\!\big|\,L^\prime \big> \times
\ \big<\alpha j_\alpha \,\big|\!\big|\, {\bf Q}_\lambda(R,\zeta) \,\big|\!\big|\,\alpha^\prime  j_{\alpha^\prime} \big> .
\label{aux1}
\end{multline}
The first reduced matrix-element is (Eq.~(4.17) of Ref.~\cite{BrS94})
\begin{multline}
\big< L \,\big|\!\big|\, {\bf Y}_\lambda(\hat{\bf R}) \,\big|\!\big|\,L^\prime \big> = \\
(-)^{\lambda - \sc L^\prime}\ \frac{\hat{\lambda}\,\hat{L}^\prime}{\sqrt{4\pi}} \ 
\left(
\begin{array}
[c]{ccc}%
\lambda\, &L^\prime & L\\
0  & 0 & 0
\end{array} 
\right).
\label{redY}
\end{multline}
Using this result, Eq.~(\ref{aux1}) can be put in the form,
\begin{equation}
X^{{\scr (i)\lambda}}_{\alpha L,\alpha^\prime L^\prime}(R)  = \mathcal{A}^{J \lambda}_{\alpha L;  \alpha^\prime L^\prime} \times F^{{\scr (i)\lambda}}_{\alpha \alpha^\prime}(R) ,
\label{X2}
\end{equation}
and Eq.~(\ref{TFi-3}) becomes
\begin{multline}
P^{\scr (i)}(J) = \frac{4K}{E\,\hat{j}_0^2}\ \sum_\lambda \, \sum_{\alpha L \alpha^\prime L^\prime\,  L_0 }\, \mathcal{A}^{J \lambda}_{\alpha L;  \alpha^\prime L^\prime} \
(i)^{\scr L^\prime - L}\\\
\times  \int dR \ {\mathcal U}^{J*}_{\alpha  L, 0  L_0}(K_{\alpha} R) \ {\mathcal U}^{J}_{\alpha^\prime  L, 0  L_0^\prime}(K_{\alpha^\prime} R)\ 
F^{{\scr (i)\lambda}}_{\alpha \alpha^\prime}(R) ,
\label{TFi-4}
\end{multline}
with
\begin{multline}
\mathcal{A}^{J \lambda}_{\alpha L;  \alpha^\prime L^\prime} = \sqrt{4\pi}\ \ (-)^{\sc J + \lambda-L-L^\prime- j_{\alpha^\prime}} \ \hat{L}\ \hat{L}^\prime\  \hat{\lambda}\ \hat{j}_{\alpha} \\
\times 
W(L L^\prime j_\alpha j_{\alpha^\prime} ;\lambda J)\ \
\left(
\begin{array}
[c]{ccc}%
\lambda\, &L^\prime & L\\
0  & 0 & 0
\end{array} 
\right)
\label{CC}
\end{multline}
and 
\begin{equation}
F^{{\scr (i)\lambda}}_{\alpha \alpha^\prime}(R) = \big<\alpha j_\alpha \,\big|\!\big|\, {\bf Q}_\lambda(R,\zeta) \,\big|\!\big|\,\alpha^\prime  j_{\alpha^\prime} \big> .
\label{FF}
\end{equation}
%

\subsection{Calculation of $F^{{\scr (i)\lambda}}_{\alpha \alpha^\prime}(R)$ for a two-fragment projectile}\label{A1-2}


Now we consider the situation where the projectile is formed by two fragments, one with spin zero and the other with spin $s$. In this case the intrinsic coordinates 
are $\zeta \equiv \{r,\hat{\bf r}\}$. The angular momentum-projected intrinsic states are then given by
\begin{equation}
\phi_{\alpha j_\alpha \nu}({\bf r}) = \frac{u_{\scr \alpha l_\alpha  j_\alpha} (r)}{r}\ \mathcal{J}_{\scr l_\alpha  j_\alpha \nu}({\bf\hat  r})  ,
\label{h}
\end{equation}
with 
\begin{equation}
\mathcal{J}_{l_\alpha j_\alpha   \nu}(\hat{\bf r})= \sum_{m_{\sc l},m_{\sc s}}
\big< l_\alpha\, m_{\sc l}\, s\, m_{ s}   \big| j_\alpha \nu  \big> \ Y_{l_\alpha m_l}(\hat{{\bf r}})\   
\left| s m_s \right>,
\end{equation}
where $\left| s m_{\scr s} \right>$ are states in the spin-space and $\big< l_\alpha\, m_{\sc l}\, s\, m_{ s}   \big| j_\alpha \nu  \big> $ are Clebsh-Gordan coefficients.
In Eq.~(\ref{h}), $u_{\scr \alpha l_\alpha  j_\alpha} (r)$ stands for the radial wave functions of the projectile. They are either bound states, or bins
generated by scattering states of the fragments, $u_{\scr \varepsilon l_\alpha  j_\alpha} (r)$, where $\varepsilon$ is the collision energy.\\

The tensor of Eq.~(\ref{expW-1-b}) then becomes
\begin{equation}
Q^{{\scr (i)}}_{\lambda \,-\mu}({\bf r}^\prime_i) \rightarrow  Q^{{\scr (i)}}_{\lambda \,-\mu}({\bf r}) = 
{\mathcal W}^{\sc (i)\lambda}(R,r)\  Y_{\lambda \, -\mu}(\hat{\bf r}) ,
\label{expW-1-d}
\end{equation}
and scalar products in the intrinsic space are integrals over $r^2 dr\ d\Omega_{\hat{\bf r}}$. \\

Then, adopting the notation of Ref.~\cite{BrS94}, Eq.~(\ref{FF}) becomes
\begin{equation}
F^{\scr (i)\lambda}_{\alpha \alpha^\prime}(R) = \mathcal{F}^{\scr (i)\lambda}_{\alpha \alpha^\prime}(R)\ \big\langle\  j_\alpha \,\big|\!\big|\, {\bf Y}_\lambda(\hat{\bf r}) \,\big|\!\big|\,  j_{\alpha^\prime} \big\rangle.
\label{FF2}
\end{equation}
where $ \mathcal{F}^{\scr (i)\lambda}_{\alpha \alpha^\prime}(R)$ is the form factor
\begin{equation}
 \mathcal{F}^{\scr (i)\lambda}_{\alpha \alpha^\prime}(R) = \int dr\ u^*_{\alpha l_\alpha j_\alpha}(r)\
  {\mathcal W}^{\sc (i)\lambda}(R,r)\ u_{\alpha^\prime l_{\alpha^\prime} j_{\alpha^\prime}}(r).
\label{FF3}
\end{equation}
The reduced matrix-element of Eq.~(\ref{FF2}) can be evaluated with help of Eq.~(5.10) of Ref.~\cite{BrS94}, and one gets
\begin{multline}
\big< j_\alpha \,\big|\!\big|\, {\bf Y}_\lambda(\hat{\bf r}) \,\big|\!\big|\, j_{\alpha^\prime} \big> 
= (-)^{\sc j_\alpha - \lambda -s+l_{\alpha^\prime}}\ \hat{l}_\alpha\,\hat{j}_{\alpha^\prime}\\
W(l_\alpha l_{\alpha^\prime} j_\alpha j_{\alpha^\prime}; \lambda s) \ \big< l_\alpha \,\big|\!\big|\, {\bf Y}_\lambda(\hat{\bf r}) \,
\big|\!\big|\, l_{\alpha^\prime} \big> .\nonumber
\end{multline}
Finally, evaluating $\big< l_\alpha \,\big|\!\big|\, {\bf Y}_\lambda(\hat{\bf r}) \,\big|\!\big|\, l_{\alpha^\prime} \big>$ as in Eq.~(\ref{redY}), the above equation becomes
\begin{multline}
\big< j_\alpha \,\big|\!\big|\, {\bf Y}_\lambda(\hat{\bf r}) \,\big|\!\big|\, j_{\alpha^\prime} \big> 
=      
 (-)^{\sc j_\alpha -s }\ \ \frac{\hat{l}_{\alpha^\prime}\,\hat{j}_{\alpha^\prime}\,\hat{l}_\alpha\,\hat{\lambda}}{\sqrt{4\pi}}\\
\left(
\begin{array}
[c]{ccc}%
\lambda\, &l_{\alpha^\prime} & l_\alpha\\
0  & 0 & 0
\end{array} 
\right)\ 
W(l_\alpha l_{\alpha^\prime} j_\alpha j_{\alpha^\prime}; \lambda s)                .
\label{redmat2}
\end{multline}
Using the above equation in Eq.~(\ref{FF2}) and inserting the result into Eq.~(\ref{TFi-4}), the fusion probability becomes
\begin{multline}
 {\mathcal P}^{\scr (i)}(J) =  \frac{4\,K}{E\,\hat{j}_0^2}\  \sum_{L_0}\ \sum_\lambda\\ 
\times  \sum_{\alpha L  \alpha^\prime L^\prime}\ \mathcal{B}^{\scr \lambda}_{\alpha L, \alpha^\prime L^\prime}(J ) \  
{\mathcal M}^{\scr (i)\lambda}_{\alpha L , \alpha^\prime L^\prime }(L_0,J).
\label{TL-6}
\end{multline}
Above, $\mathcal{B}^{\scr \lambda}_{\alpha L, \alpha^\prime L^\prime}(J )$ is the geometric factor
\begin{equation}
{\mathcal B}^{\scr \lambda}_{\alpha L, \alpha^\prime L^\prime}(J )  = \mathcal{A}^{J \lambda}_{\alpha L;  \alpha^\prime L^\prime} \ 
\big< j_\alpha \,\big|\!\big|\, {\bf Y}_\lambda(\hat{\bf r}) \,\big|\!\big|\,  j_{\alpha^\prime} \big>,
\end{equation}
or explicitly,
\begin{multline}
{\mathcal B}^{\scr \lambda}_{\alpha L, \alpha^\prime L^\prime}(J )   =  (-)^\mathcal{N}
\
\hat {\lambda}^2\,\hat{L}\,\hat{L}^\prime\,\hat{l}_\alpha\,\hat{l}_{\alpha^\prime}\,
\hat{j}_\alpha\,\hat{j}_{\alpha^\prime}  \\
\times
W\left(L L^\prime j_\alpha j_{\alpha^\prime} ; \lambda J  \right)\ 
W\left(l_\alpha l_{\alpha^\prime} j_\alpha  j_{\alpha^\prime}; \lambda s \right) \\
\times
\left(
\begin{array}
[c]{ccc}%
\lambda\, &L^\prime & L\\
0  & 0 & 0
\end{array} 
\right)
\left(
\begin{array}
[c]{ccc}%
\lambda\, &l_{\alpha^\prime} & l_\alpha\\
0  & 0 & 0
\end{array} 
\right) ,
\end{multline}
with
\begin{equation}
\mathcal{N} = J-s +j_\alpha - j_{\alpha^\prime} -L- L^\prime+\lambda,
\end{equation}
and ${\mathcal M}^{\scr (i)\lambda}_{\alpha L , \alpha^\prime L^\prime }(L_0,J)$ is the radial integral
\begin{multline}
{\mathcal M}^{\scr (i)\lambda}_{\alpha L , \alpha^\prime L^\prime }(L_0,J) =\   i^{\scr L^\prime - L}\ 
 \int dR\  {\mathcal F}^{\sc (i)\lambda}_{\alpha \alpha^\prime}(R) \\
\times {\mathcal U}^{{\sc J}*}_{\scr \alpha L,0 L_0}(K_\alpha,R)\ \,
\, \
 {\mathcal U}^{\sc J}_{\sc \alpha^\prime L^\prime,0 L_0}(K_\alpha^\prime,R).
\label{factor M}
\end{multline}

\bigskip

Although the  radial integrals are complex functions,  the probabilities of Eq.~(\ref{TL-6}) are real. 
Using symmetry properties of the 3J and Racah coefficients (see, e.g. Ref.~\cite{BrS94}) one can easily show that 
\begin{equation}
{\mathcal B}^{\scr \lambda}_{\alpha L, \alpha^\prime L^\prime}(J )  = {\mathcal B}^{\scr \lambda}_{\alpha^\prime L^\prime, \alpha L}(J ) .
\label{B-prop}
\end{equation}

\noindent On the other hand, the radial integrals have the property
\begin{equation}
{\mathcal M}^{\scr (i)\lambda}_{\alpha L , \alpha^\prime L^\prime }(L_0,J) =  
{\mathcal M}^{\scr (i)\lambda *}_{\alpha^\prime L^\prime, \alpha L}(L_0,J).
\label{M-prop}
\end{equation}
Since $\alpha,L,\alpha^\prime L^\prime$ are dummy indices running over the same ranges, the fusion probability of Eq.~(\ref{TL-6})
does not change if one interchanges $\{\alpha,L\} \rightleftarrows \{\alpha^\prime,L^\prime\}$. Then, using Eqs.~(\ref{B-prop}) and (\ref{M-prop}), one
obtains the explicitly real expression for the absorption probabilities,
\begin{multline}
 {\mathcal P}^{\scr (i)}(J) =    \frac{4\,K}{E\,\hat{j}_0^2}\  \sum_{\lambda L_0}\  \sum_{\alpha L  \alpha^\prime L^\prime}\\
 \times  \ 
B^{\scr \lambda}_{\alpha L, \alpha^\prime L^\prime}(J )\ \ \rm{Re} \Big\{ {\mathcal M}^{\scr (i)\lambda}_{\alpha L , \alpha^\prime L^\prime }(L_0,J)
\Big\} .
\label{TL-76}
\end{multline}
Finally, the probabilities ${\mathcal P}^{\scr (i)}_{\scr B}(J) $ and ${\mathcal P}^{\scr (i)}_{\scr C}(J) $ are given by the above expression, restricting the
sum over channels to $\{ \alpha,\alpha^\prime \} \in B$ and to $\{\alpha,\alpha^\prime \}\in C$, respectively.\\

\bigskip

\centerline{ ACKNOWLEDGEMENTS}

\medskip

Work supported in part by the Brazilian funding agencies, CNPq, FAPERJ, and the INCT-FNA (Instituto Nacional de Ci\^encia e 
Tecnologia- F\'\i sica Nuclear e Aplica\c c\~oes), research project 464898/2014-5. We are indebted to professor Raul Donangelo for critically reading the manuscript.



\end{document}